\documentclass[journal]{IEEEtran}
\usepackage{booktabs}
\usepackage{graphicx}  
\usepackage{epstopdf}
\usepackage{comment}
\usepackage{multirow}
\usepackage{enumerate}
\usepackage{color}
\usepackage[linewidth=1pt]{mdframed}
\usepackage{float}
\usepackage{amsmath,amsfonts}
\usepackage{algorithmic}
\usepackage{array}
\usepackage[caption=false,font=normalsize,labelfont=sf,textfont=sf]{subfig}
\usepackage{textcomp}
\usepackage{stfloats}
\usepackage{url}
\usepackage{verbatim}
\usepackage{cuted}
\def\BibTeX{{\rm B\kern-.05em{\sc i\kern-.025em b}\kern-.08em
    T\kern-.1667em\lower.7ex\hbox{E}\kern-.125emX}}
\usepackage{balance}

\begin{document}

\title{Modeling of UV NLoS Communication Channels: From Atmospheric Scattering and Obstacle\\ Reflection Perspectives}

\author{Tianfeng~Wu,~\IEEEmembership{Graduate Student Member, IEEE,}
            Fang~Yang,~\IEEEmembership{Senior Member,~IEEE,}
            Tian~Cao,~\IEEEmembership{Member,~IEEE,}
            Ling~Cheng,~\IEEEmembership{Senior Member,~IEEE,}
            Yupeng~Chen,
	    Jian~Song, \IEEEmembership{Fellow,~IEEE,}\\
            Julian~Cheng,~\IEEEmembership{Fellow,~IEEE,}
            and~Zhu~Han,~\IEEEmembership{Fellow,~IEEE}

\thanks{Part of this paper has been accepted by IEEE GLOBECOM, 2024~\cite{ref1}. This work was supported by the National Key Research and Development Program of China (2023YFE0110600) and was partially supported by NSF ECCS-2302469, Toyota. Amazon and Japan Science and Technology Agency (JST) Adopting Sustainable Partnerships for Innovative Research Ecosystem (ASPIRE) JPMJAP2326. (\emph{Corresponding author: Fang Yang}).} 
\thanks{Tianfeng Wu and Fang Yang are with the Department of Electronic Engineering, Beijing National Research Center for Information Science and Technology (BNRist), Tsinghua University, Beijing 100084, P. R. China (e-mail: wtf22@mails.tsinghua.edu.cn; fangyang@tsinghua.edu.cn).}
\thanks{Tian Cao is with the School of Telecommunications Engineering, Xidian University, Xi’an 710071, P. R. China (e-mail: caotian@xidian.edu.cn).}
\thanks{Ling Cheng is with the School of Electrical and Information Engineering, University of the Witwatersrand, Johannesburg 2000, South Africa (email: ling.cheng@wits.ac.za).}
\thanks{Yupeng Chen is with the School of Microelectronics, Southern University of Science and Technology, Shenzhen 518055, P. R. China (email: chenyp@sustech.edu.cn).}
\thanks{Jian Song is with the Department of Electronic Engineering, Tsinghua University, Beijing 100084, P. R. China, and also with the Shenzhen International Graduate School, Tsinghua University, Shenzhen 518055, P. R. China (e-mail: jsong@tsinghua.edu.cn).}
\thanks{Julian Cheng is with the School of Engineering, The University of British Columbia, Kelowna, BC, V1V 1V7, Canada (e-mail: julian.cheng@ubc.ca).}
\thanks{Zhu Han is with the Department of Electrical and Computer Engineering at the University of Houston, Houston, TX 77004 USA, and also with the Department of Computer Science and Engineering, Kyung Hee University, Seoul, South Korea, 446-701 (e-mail: hanzhu22@gmail.com).}}

\markboth{IEEE Journal on Selected Areas in Communications}%
{How to Use the IEEEtran \LaTeX \ Templates} 

\maketitle

\begin{abstract}
As transceiver elevation angles increase from small to large, existing ultraviolet (UV) non-line-of-sight (NLoS) models encounter two challenges: i) cannot estimate the channel characteristics of UV NLoS communication scenarios when there exists an obstacle in the overlap volume between the transmitter beam and the receiver field-of-view (FoV), and ii) cannot evaluate the channel path loss for the wide beam and wide FoV scenarios with existing simplified single-scattering path loss models. To address these challenges, a UV NLoS scattering model incorporating an obstacle was investigated, where the obstacle's orientation angle, coordinates, and geometric dimensions were taken into account to approach actual application environments. Then, a UV NLoS reflection model was developed combined with specific geometric diagrams. Further, a simplified single-scattering path loss model was proposed with a closed-form expression. Finally, the proposed models were validated by comparing them with the Monte-Carlo photon-tracing model, the exact single-scattering model, and the latest simplified single-scattering model. Numerical results show that the path loss curves obtained by the proposed models agree well with those attained by related NLoS models under identical parameter settings, and avoiding obstacles is not always a good option for UV NLoS communications. Moreover, the accuracy of the proposed simplified model is superior to that of the existing simplified model for all kinds of transceiver FoV angles.                                                                                                         
\end{abstract}


\IEEEpeerreviewmaketitle          

\section{Introduction} 
\IEEEPARstart{N}{owadays}, radio frequency (RF) technology is widely adopted in wireless communication scenarios due to its attractive benefits,  for example, low cost and easy deployment. However, this technology also encounters certain issues such as spectrum scarcity, mutual interference, and communication security \cite{ref2,ref3}. To solve these problems, much research work has been conducted on millimeter-wave (mm-Wave), terahertz (THz), infrared (IR), and visible light communications (VLC) \cite{ref4,ref5}. Mm-Wave and THz communications deploy a lot of antenna arrays with high beamforming gains to overcome the severe link loss \cite{ref4}. IR and VLC convey information based on the existing optical transceivers and illumination networks~\cite{ref6,ref7}. However, mm-Wave, THz, IR, and VLC primarily depend on line-of-sight (LoS) links for communications and are thus susceptible to obstacles. 

To address the issues above and promote the development of the next-generation wireless communication technology, much effort has been devoted to the investigation of ultraviolet (UV) communications \cite{ref2,ref5,ref8}--\cite{ref11}. Compared with RF communications, UV communications have inherent advantages such as huge bandwidth without licensing, high confidentiality, and immunity to electromagnetic interference \cite{ref5,ref9}. Compared with mm-Wave, THz, IR, and VLC, UV can convey information via non-line-of-sight (NLoS) links that utilize atmospheric scattering \cite{ref12}. Moreover, the background noise of the UV communication systems adopting 200-280 nm band (also called the solar-blind band) can be neglected. Since the atmospheric ozone layer has a strong absorption impact on the solar radiation within this band \cite{ref13}. Encouraged by these benefits, UV communication technology has great potential to enable future ubiquitous communications by integrating with existing wireless communication technologies.   

\linespread{1.5}
\begin{table*}
\centering
\linespread{1.0}
\caption{Applicability of Representative UV NLoS Channel Models}
\begin{tabular}{|c|c|c|l|c|}
\hline
{\textbf{Types}} & {\textbf{Refs}} & {\textbf{Obstacle}} & \multicolumn{1}{c|}{\textbf{Applicability}} & \textbf{Range} \\ 
\hline
\multirow{5}{*}{SSM}&{\cite{ref12,ref21}}&{$\times$}&{The transceiver points arbitrarily in the coplanar case.}&{Short}\\
\cline{2-5}
{}&{\cite{ref18,ref22}}&{$\times$}&{The transceiver points arbitrarily in both coplanar and non-coplanar cases.}&{Short}\\
\cline{2-5}
{}&{\cite{ref19}}&{$\times$}&{Vertical receiver pointing and arbitrary transmitter orientation.}&{Short}\\
\cline{2-5}
{}&{\cite{ref20}}&{$\surd$}&{Arbitrary transceiver pointing in both coplanar and non-coplanar cases and an infinite reflection plane.}&{Short}\\
\cline{2-5}
{}&{\cite{ref23}}&{$\times$}&{Transceiver FoVs are above the horizontal plane in the coplanar case with a close overlap volume.}&{Short}\\
\hline
\multirow{4}{*}{SSSPLM}&{\cite{ref24}}&{$\times$}&{A small overlap volume between transceiver FoVs in the coplanar case.}&{Short}\\
\cline{2-5}
{}&{\cite{ref25}}&{$\times$}&{A small overlap volume between transceiver FoVs in both coplanar and non-coplanar cases.}&{Short}\\
\cline{2-5}
{}&{\cite{ref26}}&{$\times$}&{A narrow transmitter beam in both coplanar and non-coplanar cases.}&{Short}\\
\cline{2-5}
{}&{\cite{ref27,ref28}}&{$\times$}&{A narrow transmitter beam or receiver FoV in both coplanar and non-coplanar cases.}&{Short}\\
\hline
\multirow{3}{*}{MSM}&{\cite{ref29}--\cite{ref31}}&{$\times$}&{The transceiver points arbitrarily in both coplanar and non-coplanar cases.} & {Medium}\\
\cline{2-5}
{}&{\cite{ref32}}&{$\surd$}&{Arbitrary transceiver pointing in both coplanar and non-coplanar cases and an infinite reflection plane.}&{Medium}\\
\cline{2-5}
{}&{\cite{ref33}}&{$\surd$}&{Arbitrary transceiver pointing in the coplanar case and considering an obstacle's width and height.}&{Medium}\\
\hline
\multirow{2}{*}{SMSM}&{\cite{ref34}}&{$\times$}&{The transceiver points arbitrarily in both close and open overlap volume cases.} & {Medium}\\
\cline{2-5}
{}&{\cite{ref35}}&{$\times$}&{The contributions of second-order and third-order scattering to the entire signal in a small beam case.}&{Medium}\\
\hline
\multirow{3}{*}{TCM}&{\cite{ref36}}&{$\times$}&{The transceiver points arbitrarily in weak-medium turbulence conditions.} & {Long}\\
\cline{2-5}
{}&{\cite{ref37,ref38}}&{$\times$}&{A small overlap volume between transceiver FoVs in weak-medium turbulence conditions.}&{Long}\\
\cline{2-5}
{}&{\cite{ref40}}&{$\times$}&{A narrow transmitter beam in weak-medium turbulence conditions.}&{Long}\\
\hline
\multirow{2}{*}{EPLM}&{\cite{ref43}}&{$\times$}&{Some fixed transceiver geometries in the coplanar case.} & {Short}\\
\cline{2-5}
{}&{\cite{ref44}}&{$\times$}&{Some fixed transceiver geometries in both coplanar and non-coplanar cases.}&{Short}\\
\hline
\end{tabular}
\label{t1}
\end{table*}
\linespread{1.0}  
The current challenges to UV NLoS communications can be briefly summarized as follows. Concerning hardware systems, the power efficiencies of all current UV light sources are still low, and the optical detectors loaded with solar blocking filters are expensive \cite{ref2,ref3}, which need to be tackled to expand the application range of UV communication technology. Besides, UV NLoS systems suffer from high channel fading and severe pulse-broadening effects \cite{ref14}, which require more efficacious modulation and coding schemes to complete the tasks of high-speed data transmission. Moreover, given the high complexity of actual communication environments, more accurate channel models incorporating factors such as diversified scenarios with obstacles, joint effects of atmospheric turbulence and multiple-scattering events, various weather conditions, and time-variant channel parameters should be developed to estimate UV NLoS propagation characteristics \cite{ref15,ref16}. Furthermore, the signals received by the UV detector may be mixed with various noises such as optical background noise, thermal noise, and detector dark noise \cite{ref2}, so advanced detection techniques are required to extract signals overwhelmed by these noises. In addition to that, in UV network communications, medium access control (MAC) and multi-user interference (MUI) seriously influence network performance. How to design efficient MAC protocols and eliminate MUI is challenging for the development of UV network communications owing to rich atmospheric scattering. In this paper, we investigate the channel models for UV NLoS scenarios incorporating an obstacle and the simplified channel model for UV NLoS scenarios without any obstacle.

\subsection{Prior Works}
At present, research on UV communication channel models has primarily focused on the interactions between UV photons and atmospheric constituents, e.g., molecules and aerosols. To be specific, when the communication range is small and there is an overlap volume of the transceiver field-of-views (FoVs), UV signals received by the receiver (R) are dominated by the single-scattering events \cite{ref17}. Therefore, many single-scattering models (SSMs) were proposed to obtain the channel properties of short-range UV NLoS communication scenarios, including the channel models developed based on the prolate-spheroidal coordinate system \cite{ref12,ref18} and the channel models proposed based on the spherical coordinate system \cite{ref19}--\cite{ref23}, where the model in \cite{ref20} considered the obstacle reflection, and idealized the obstacle as an infinite plane. Nevertheless, due to the high complexity of the triple integral limits in \cite{ref12}, \cite{ref18}--\cite{ref23}, some simplified single-scattering path loss models (SSSPLMs) were put forward for tractable analysis \cite{ref24}--\cite{ref28}, where the model proposed in \cite{ref24} applies to the case where the transceiver FoV axes are on a common plane, while the rest simplified models are suitable for the cases where the transceiver FoV axes are both coplanar and non-coplanar. 

As the overlap volume between the transmitter (T) beam~and receiver FoV decreases or the communication range increases, the accuracy of the single-scattering model decreases gradually \cite{ref21}, and the contribution of multiple-scattering events to the entire received signal becomes significant \cite{ref29}. Hence, several multiple-scattering models (MSMs) were developed based on the Monte-Carlo method \cite{ref29}--\cite{ref33}, where numerous photons are required to participate in the simulation to obtain reliable signal statistics at the receiver. In \cite{ref29}, the impacts of various system parameters on channel path loss and bandwidth were investigated. With respect to \cite{ref30}, the contribution of different scattering orders to path loss and impulse response functions was analyzed. Also, the relevant pseudocode was provided to reproduce simulation results. Regarding \cite{ref31}, the convergence performance of the Monte-Carlo integration model was studied and improved by introducing different sampling methods, and concerning \cite{ref32}, an infinite reflection plane was introduced to improve received optical power. As regards \cite{ref33}, the obstacle's height and width were taken into account, while its thickness was assumed to be infinitely large. However, owing to the high computational complexity of these Monte-Carlo models, some simplified MSMs (SMSMs) were derived to simplify analysis, including a sample-based UV channel characterization method that incorporates first-order and second-order scattering events \cite{ref34}, and an integral model that incorporates second-order and third-order scattering events \cite{ref35}. Moreover, several turbulence channel models (TCMs) \cite{ref36}--\cite{ref42} were proposed to assess the channel properties of long-range UV NLoS communications, since the turbulence impact cannot be ignored when the range exceeds 200 m \cite{ref36}. In addition to these works, two empirical path loss models (EPLMs) including the coplanar model \cite{ref43} and the non-coplanar model \cite{ref44} were developed by fitting the extensive outdoor measurements. For clarity, the applicability of the aforementioned UV NLoS models is provided in Table~\ref{t1}, where the communication ranges related to the terms ``short", ``medium'', and ``long'' can be summarized as less than 100~m, less than 200~m, and greater than 200 m, respectively, based on the existing works \cite{ref23,ref29,ref30,ref36,ref43}, and the terms ``$\surd\,$'' and ``$\times$'' represent the models with and without obstacles, respectively.

Although tremendous efforts have been implemented on UV NLoS communications, channel modeling is still an open issue owing to the complexity of application environments \cite{ref2,ref13}. According to \cite{ref2,ref32}, it can be found that the existing studies on UV NLoS channel modeling primarily focus on scenarios without obstacles, which makes them unsuitable for the small transceiver elevation angles in most cases, e.g., scenarios with multiple buildings or mountainous regions. Specifically, when transceiver elevation angles are small, obstacles will inevitably appear in their FoVs and affect the propagation process of UV signals. Regarding the models in \cite{ref20,ref32,ref33}, they idealize the obstacle as an infinite plane (considering obstacle reflection and partial obstruction of signal transmission links) or suppose that the obstacle thickness is infinitely large (only the blocking of signaling links by the obstacle is considered), which greatly limits their applicabilities. Concerning the existing SSSPLMs, which are summarized in Table~\ref{t1}, they only apply to the cases where the overlap volume between transceiver FoVs is small, or the transmitter beam angle (or receiver FoV angle) is small. This is because in these cases, the scattering characteristics of the points in the overlap volume can be assumed to be the same as that of the intersection point between transceiver FoV axes, or to be consistent with that of the scattering points located on the transceiver FoV axes, whereas when the overlap volume~or transceiver FoVs is relatively large, these assumptions can lead to a large error.

\subsection{Contributions of This Paper}
To solve the aforementioned problems, we comprehensively investigate the channel models for short-range UV NLoS communication scenarios in this paper. In regard to scenarios with obstacles, the total received pulse energy consists of two parts: i) the energy contributed by atmospheric scattering, and ii) the energy contributed by the obstacle reflection. To facilitate the understanding, we model the two types of energies separately. For scenarios without obstacles, we propose a simplified path loss model with a closed-form expression. Compared with the related works \cite{ref20,ref32,ref33}, the proposed integration model applies to the cases with arbitrary obstacle parameter settings, where the obstacle is modeled as the cuboid and accords with most practical application scenarios in urban areas. Compared with the typical SSSPLMs \cite{ref24}--\cite{ref28}, the proposed simplified path loss model applies to the cases with diversified transceiver FoV angles. For clarity, the main contributions of the paper~are summarized as follows: 
\begin{itemize}
\item A UV NLoS scattering model incorporating an obstacle is investigated, where the obstacle's coordinates, thickness, width, height, and orientation are considered to approach real communication environments\footnote{To make the proposed model easy to understand, the whole derivation process is combined with detailed cases, scenarios, conditions, and a flow chart.}. Through this model, the energy contribution of atmospheric scattering to the signal can be obtained. 
\item A UV NLoS reflection model is developed in combination with the projection of the obstacle on the horizontal plane, which can ensure that all possible reflection surfaces are not missed. In this part, the reflection pattern of the plane is modeled as the Phong model since its effectiveness has been experimentally verified. With the help of this model, the energy contribution of obstacle reflection to the signal can be acquired.
\item Owing to the high computational complexity of the exact SSMs without obstacles, whose upper and lower limits on triple integrals are complex expressions, and the restricted applicabilities of the existing SSSPLMs, which apply to the communication scenarios with small transceiver FoV angles or small overlap volume, an SSSPLM is proposed with a closed-form expression. First, the transmitter beam is sampled uniformly into multiple sub-beams, which can lower the computational complexity of the path loss from the triple integral to the single integral. Second, the single integral corresponding to each sub-beam is eliminated via the Gauss-Legendre quadrature rule. Finally, the path loss expression is obtained by superimposing the contribution of each sub-beam to the signal.     
\item The proposed NLoS models are validated by comparing them with the relevant works. Calculation and simulation results show that the proposed models coincide well with the existing works under identical parameter settings, and bypassing obstacles is not always a good option for UV NLoS communications. Beyond that, the accuracy of the proposed SSSPLM is superior to that of the state-of-the-art SSSPLM for arbitrary transceiver FoV angles. 
\end{itemize} 

The remainder of the paper is organized as follows. First,~the system model for UV NLoS communications is elaborated in Section II. Following that, the UV NLoS scattering model and reflection model incorporating an obstacle are investigated in Section III and Section IV, respectively. Moreover, a simplified single-scattering path loss model is proposed in Section V with a closed-form expression. Further, the proposed NLoS models are examined in Section VI. Finally, conclusions are drawn in Section VII. 

\textit{Notations}: Throughout this paper, boldface lowercase letters (e.g. $\boldsymbol{n}$) and boldface uppercase letter pairs (e.g. $\boldsymbol{\rm{TF}}$) represent vectors, and $||\cdot||$ denotes the Euclidean norm of a vector.

\begin{figure}[t]  
\centering  
\includegraphics[scale=0.366]{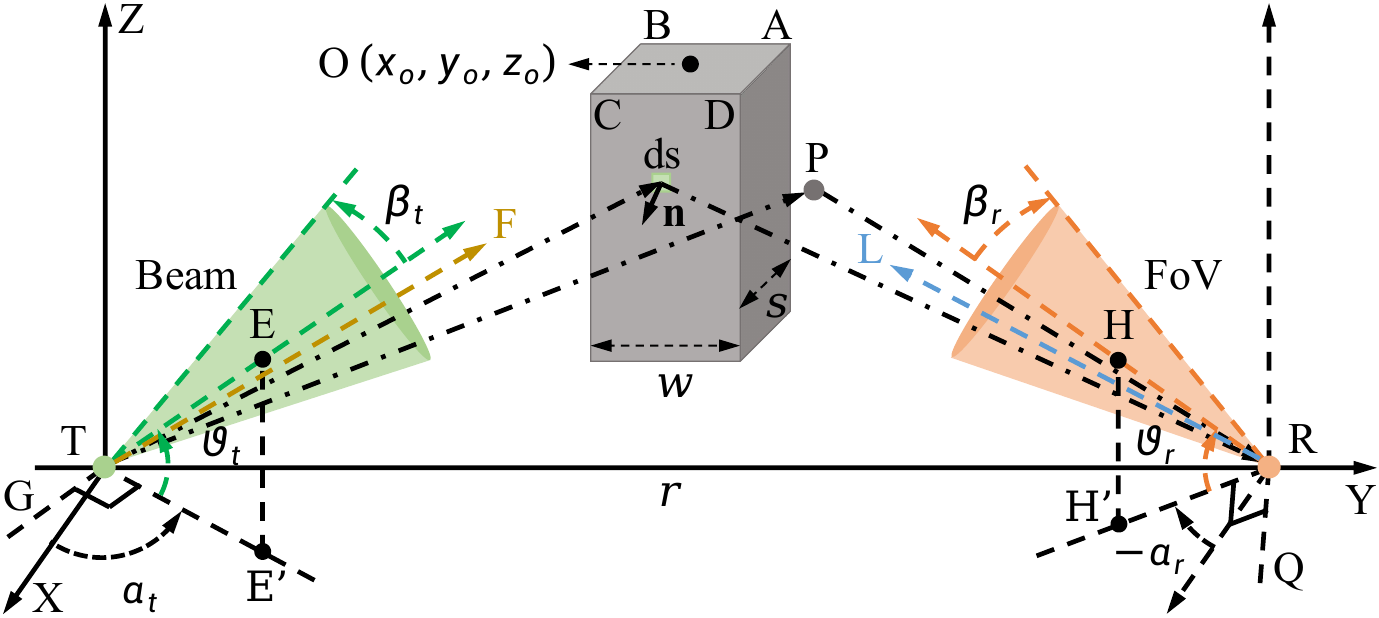}
\centering
\caption{Illustration of UV NLoS communication links with an obstacle.}
\label{Fig1}  
\end{figure}
\section{System Model}
\begin{figure}[t]  
\centering  
\includegraphics[scale=0.40]{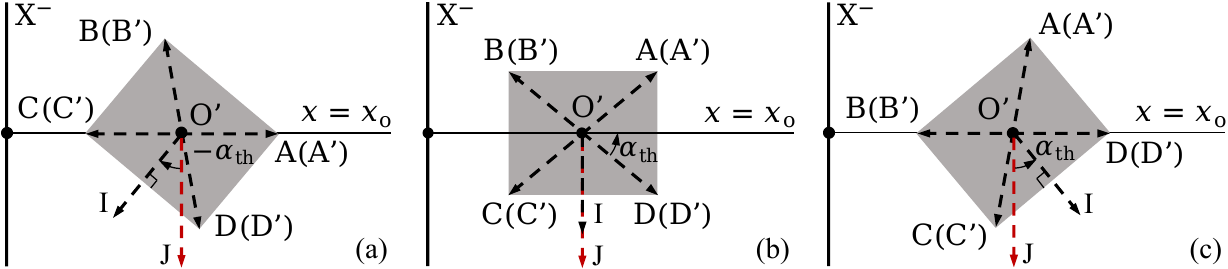}
\centering
\caption{The projection of the obstacle on the plane $\rm{X^{-}Y}$, where $\alpha=-\alpha_{\rm{th}}$ for (a), $\alpha=0$ for (b), and $\alpha=\alpha_{\rm{th}}$ for (c).}
\label{Fig2}  
\end{figure}

As shown in Fig.~\ref{Fig1}, the parameters of the system model are defined as follows: $\beta_t$ and $\beta_r$ denote the half-beam angle and half-FoV angle of the transmitter and receiver, respectively; $\vartheta_t$ is the transmitter elevation angle and is positive if taken anti-clockwise from $\rm{TE'}$, where $\rm{E'}$ is the projection of the point E on the plane XY, and this representation also applies to other points such as A and B; $\vartheta_r$ is the receiver elevation angle and which is positive if taken clockwise from $\rm{RH'}$, where rays TE and RH are the transmitter beam axis and receiver FoV axis, respectively; $\alpha_t$ and $\alpha_r$ denote the azimuth angles of T and R, respectively, which are positive if taken anticlockwise from X positive axis; $\tau$ and $\varepsilon$ denote the distances from the scattering point P (or the reflection region d$\mathbb{U}$) to T and R, respectively; $\vartheta_s$ and $\vartheta_v$ are the angles between $\boldsymbol{\rm{TP}}$ and $\boldsymbol{\rm{PR}}$, and $\boldsymbol{\rm{RP}}$ and $\boldsymbol{\rm{RH}}$, respectively; $A_r$ represents the receiver aperture area; $r$ is the communication range; $w$ and $s$ are the obstacle’s width and thickness, respectively; $\kappa$ denotes the height of the obstacle above the plane XY; ${\rm{O}}\,(x_o, y_o, z_o)$ is the central coordinate of the obstacle; and $\alpha$ represents the orientation angle of the obstacle shown in Fig.~\ref{Fig2} and is defined as the angle between $\boldsymbol{\rm{O'J}}$ and $\boldsymbol{\rm{O'I}}$, which is positive when rotating anticlockwise from $\boldsymbol{\rm{O'J}}$ and negative otherwise.

In order to analyze the intersection circumstances between the transceiver FoVs and the obstacle, two planes and several variables need to be defined. As shown in Fig.~\ref{Fig1}, $\mathcal{F}_{\vartheta}$ is defined as the plane passing through the line TF and perpendicular to the plane $\rm{TEE'}$, which rotates around the line TG, where TG is located on the plane XY and perpendicular to the line $\rm{TE'}$. Besides, the subscript $\vartheta$ represents the angle between $\boldsymbol{\rm{TE}}$ and $\boldsymbol{\rm{TF}}$, which is positive if taken anticlockwise from $\boldsymbol{\rm{TE}}$. $\mathcal{L}_{\sigma}$ is defined as the plane passing through RL and perpendicular to the plane $\rm{RHH'}$, which rotates around the line RQ, where RQ is located on the plane XY and perpendicular to the line $\rm{RH'}$. Besides, the subscript $\sigma$ represents the angle between $\boldsymbol{\rm{RH}}$ and $\boldsymbol{\rm{RL}}$, which is positive if rotating clockwise from $\boldsymbol{\rm{RH}}$. Beyond that, ${\rm{P}}_{t, m n}$ denotes the intersection point of the plane $\mathcal{F}_{\vartheta}$ with the line MN (e.g. ${\rm{P}}_{t, a a'}$ and the line $\rm{AA'}$); $\Theta_{t,n}$ is the angle between the plane $\mathcal{F}_{-\vartheta_t}$ and the plane TGN (e.g. $\Theta_{t,b}$ and the plane TGB); $\Psi_{t, m n}$ denotes the angle between $\boldsymbol{{\rm{TP}}_{t, m n}}$ and $\boldsymbol{\rm{TK}}$ (e.g. $\Psi_{t, a a'}$ and $\boldsymbol{{\rm{TP}}_{t, a a'}}$), where TK is the intersection ray between the plane YZ and the plane $\mathcal{F}_{\vartheta}$; ${\rm{P}}_{r, m n}$ represents the intersection point of the plane $\mathcal{L}_{\sigma}$ with the line MN (e.g. ${\rm{P}}_{r, a a'}$ and the line $\rm{AA'}$); $\Theta_{r,n}$ represents the angle between the plane $\mathcal{L}_{-\vartheta_r}$ and the plane RQN (e.g. $\Theta_{r,d}$ and the plane RQD); $\Psi_{r, m n}$ represents the angle between $\boldsymbol{\rm{RS}}$ and $\boldsymbol{{\rm{RP}}_{r, m n}}$, where RS is the intersection ray between planes $\mathcal{L}_{\sigma}$ and YZ; and $\varpi$ denotes the angle between $\boldsymbol{\rm{TF}}$ and $\boldsymbol{\rm{TV}}$ ($\boldsymbol{\rm{TV}}$ is located on the plane $\mathcal{F}_{\vartheta}$ and enclosed by the transmitter beam), which is positive when rotating anticlockwise from $\boldsymbol{\rm{TF}}$. Note that M and N are dummy variables, which can denote different points.  

Moreover, the atmospheric model parameters are defined as follows based on the existing works \cite{ref24,ref30}. Specifically, $k_e$ represents the atmospheric extinction coefficient, which can be obtained by adding the scattering coefficient $k_s$ and absorption coefficient $k_a$, and $k_s$ can be acquired by adding the Rayleigh scattering coefficient $k_s^{\rm{Ray}}$ and Mie scattering coefficient $k_s^{\rm{Mie}}$. Besides, ${\rm{P}}(\cos{\vartheta_s})$ denotes the scattering phase function \cite{ref24}, which is modeled as a weighted sum of the Rayleigh scattering phase function and Mie scattering phase function, where $\gamma$, g, and $f$ are associated model parameters \cite{ref30}.

Next, we specify the intervals of some variables mentioned above. Given the symmetry of the obstacle, the interval of $\alpha$ is set to $[-\alpha_{\rm{th}},\alpha_{\rm{th}}]$, which can characterize all situations those occur during the rotation of the obstacle, where the threshold angle $\alpha_{\rm{th}}$ can be derived as
\begin{equation}
\alpha_{\rm{th}}=\tan^{-1}(s/w).
\end{equation}
Then, the intervals of $x_o$ and $y_o$ are set to $(-\infty,x_{o,\max})$ and $(y_{o,\min},r-y_{o,\min})$, respectively, where $x_{o,\max}$ and $y_{o,\min}$ can be expressed as       
\begin{subequations}
\begin{equation}
x_{o,\max}=-r_{ws}\sin(\alpha_{\rm{th}}+|\alpha|),
\end{equation}
\begin{equation}
y_{o,\min}=r_{ws}\sin({\pi}/{2}-\alpha_{\rm{th}}+|\alpha|),
\end{equation}
\end{subequations}
and $r_{ws}=(w^2+s^2)^{1/2}/2$. Besides, we set $\alpha_t$ to $[\pi/2,\pi)$, $\alpha_r$ to $(-\pi,-\pi/2]$, and $\vartheta_t$, $\vartheta_r$, $\beta_t$, and $\beta_r$ all to $(0,\pi/2)$, which satisfy the requirements of general UV NLoS communication scenarios.

Further, the intersection orders between $\mathcal{F}_{\vartheta}$ and the obstacle vertices A, B, C, and D are investigated. Based on geometric relationships, the X and Y coordinates of A, B, C, and D can be derived as
\begin{subequations}
\begin{align}
{\rm{A}}:&
\begin{cases}
x_a=-r_{ws}\sin(\alpha+\alpha_{\rm{th}})+x_o,\\
y_a=r_{ws}\cos(\alpha+\alpha_{\rm{th}})+y_o,
\end{cases}
\\{\rm{B}}:&
\begin{cases}
x_b=-r_{ws}\sin(-\alpha+\alpha_{\rm{th}})+x_o,\\
y_b=-r_{ws}\cos(-\alpha+\alpha_{\rm{th}})+y_o,
\end{cases}
\end{align}
\begin{align}
{\rm{C}}:&
\begin{cases}
x_c=r_{ws}\sin(\alpha+\alpha_{\rm{th}})+x_o,\\
y_c=-r_{ws}\cos(\alpha+\alpha_{\rm{th}})+y_o,
\end{cases}
\\{\rm{D}}:&
\begin{cases}
x_d=r_{ws}\sin(-\alpha+\alpha_{\rm{th}})+x_o,\\
y_d=r_{ws}\cos(-\alpha+\alpha_{\rm{th}})+y_o,
\end{cases}
\end{align}
\end{subequations} 
and their Z coordinates are all $\kappa$. As $\mathcal{F}_{\vartheta}$ rotates anticlockwise from $\mathcal{F}_{-\vartheta_t}$ to $\mathcal{F}_{\pi/2-\vartheta_t}$, the possible intersection orders of $\mathcal{F}_{\vartheta}$ with A, B, C, and D are summarized in Table~\ref{t2}, where $\Theta_{t,a}$, $\Theta_{t,b}$, $\Theta_{t,c}$, and $\Theta_{t,d}$ can be given by
\begin{equation}
\Theta_{t,n}=\tan^{-1}\left(\frac{\kappa\sqrt{\cot^2{\alpha_t}+1}}{|x_n\cot{\alpha_t}+y_n|}\right),
\end{equation} 
and the subscript $n$ can be $a$, $b$, $c$ or $d$.
\linespread{1.8}
\begin{table}[t]
\centering
\linespread{1.0}
\caption{Intersection Orders of the Plane $\mathcal{F}_{\vartheta}$ with A, B, C, and D}
\label{t2}
\begin{tabular}{| c | c |}
\hline
\multirow{2}{*}{\textbf{All Scenarios}}&$\boldsymbol{\Theta_{t,a}}$, $\boldsymbol{\Theta_{t,b}}$, $\boldsymbol{\Theta_{t,c}}$, \textbf{and} $\boldsymbol{\Theta_{t,d}}$\\
{}&\textbf{in descending order}\\
\hline 
{Scenario T1}&{$\Theta_{t,c}>\Theta_{t,b}>\Theta_{t,d}>\Theta_{t,a}$}\\
\hline
{Scenario T2}&{$\Theta_{t,c}>\Theta_{t,d}\ge\Theta_{t,b}>\Theta_{t,a}$}\\
\hline
{Scenario T3}&{$\Theta_{t,c}=\Theta_{t,d}>\Theta_{t,b}=\Theta_{t,a}$}\\
\hline
{Scenario T4}&{$\Theta_{t,b}>\Theta_{t,c}>\Theta_{t,a}>\Theta_{t,d}$}\\
\hline
{Scenario T5}&{$\Theta_{t,b}=\Theta_{t,c}>\Theta_{t,a}=\Theta_{t,d}$}\\
\hline
{Scenario T6}&{$\Theta_{t,d}>\Theta_{t,c}>\Theta_{t,a}>\Theta_{t,b}$}\\
\hline
\end{tabular}
\end{table}
\linespread{1.0}  
\begin{figure}[t]  
\centering  
\includegraphics[scale=0.50]{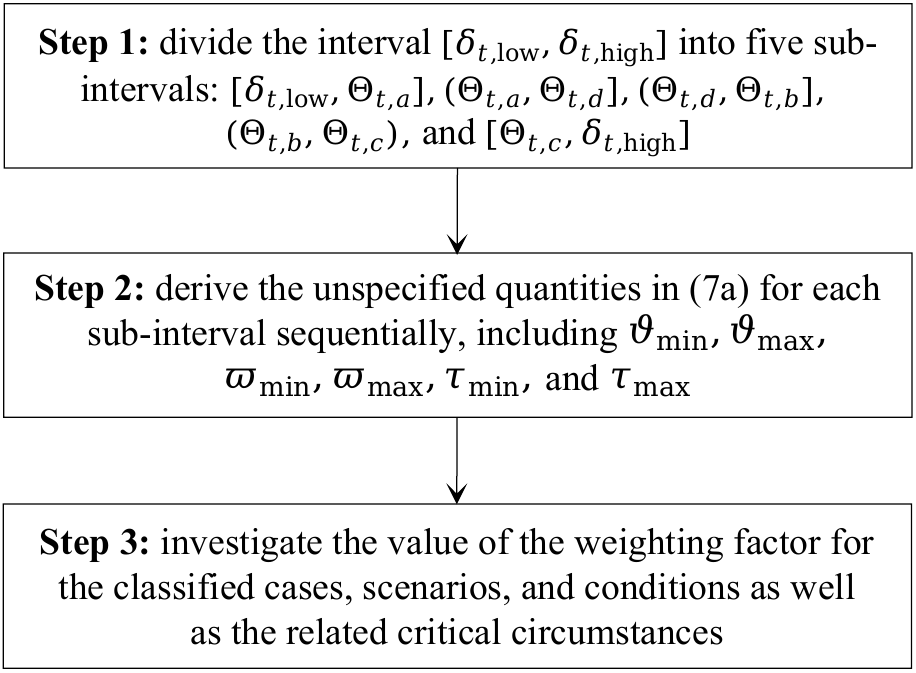}
\centering
\caption{The flow chart for solving the unspecified quantities in (\ref{eq:7a}).}
\label{Fig3}  
\end{figure} 

When the pulse energy $\mathcal{Q}_t$ is emitted by the transmitter, the composition of the received pulse energy, $\mathcal{Q}_r$, is closely related to the angle $\delta_{t,\rm{low}}=\vartheta_t-\beta_t$, which corresponds to the plane $\mathcal{F}_{-\beta_t}$. Specifically, if $\delta_{t,\rm{low}}<\Theta_{t,\rm{max}}=\max(\Theta_{t,a},\Theta_{t,b},\Theta_{t,c},$ $\Theta_{t,d})$, which implies that the transmitter beam may intersect with the obstacle, $\mathcal{Q}_r$ may compose of two parts: the energy contributed by atmospheric scattering, $\mathcal{Q}_{r,\rm{sca}}$, and the energy contributed by obstacle reflection, $\mathcal{Q}_{r,\rm{ref}}$. If $\delta_{t,\rm{low}}\ge\Theta_{t,\rm{max}}$, which means that the transmitter beam does not intersect with the obstacle, $\mathcal{Q}_r$ is only composed of the energy contributed by atmospheric scattering. In the following, we will derive the two types of energies combined with specific obstacle scenarios.

Throughout the entire modeling process, three assumptions are presented. First, we suppose that UV photons are scattered once by the atmosphere before reaching the receiver, which is consistent with the existing works \cite{ref12}, \cite{ref18}--\cite{ref23}. This means that the proposed models apply to short-range communication scenarios. Then, $w$ is assumed to be greater than $s$, which can reduce the possibility of the order in which the planes $\mathcal{F}_{\vartheta}$ and $\mathcal{L}_{\sigma}$ intersect with the obstacle vertices as well as the relevant modeling complexity. Note that this restriction can be removed based on the proposed modeling method in this paper. Further, we suppose that $\delta_{t,\rm{low}}$ and $\delta_{r,\rm{low}}=\vartheta_r-\beta_r$ are greater than $0$, which can avoid the energy loss caused by part of the UV beam pointing to the ground, and assume that $\delta_{t,\rm{high}}=\vartheta_t+$ $\beta_t$ and $\delta_{r,\rm{high}}=\vartheta_r+\beta_r$ are less than $\pi/2$, which reduces the possibility that the overlap volume is not closed and enhances the validity of the single-scattering assumption.

\section{Modeling of UV NLoS Scattering Channel Incorporating An Obstacle}
In this section, we take the representative Scenario T1 as an example to derive $\mathcal{Q}_{r,\rm{sca}}$, where the relationship among $\delta_{t,\rm{low}}$, $\delta_{t,\rm{high}}$, $\Theta_{t,a}$, $\Theta_{t,b}$, $\Theta_{t,c}$, and $\Theta_{t,d}$ must be discussed under the constraint that $\delta_{t,\rm{low}}$ is smaller than $\Theta_{t,\rm{max}}$. Considering that the situation of $\delta_{t,\rm{high}}>\Theta_{t,c}>\Theta_{t,b}>\Theta_{t,d}>\Theta_{t,a}>\delta_{t,\rm{low}}$ contains the most comprehensive possibilities, we choose it to conduct the detailed derivation. 

When $\mathcal{Q}_t$ is emitted by the transmitter, the unextinguished energy arriving at the differential volume ${\rm{d}}v$ will be scattered. Referring to \cite{ref12}, ${\rm{d}}v$ can be considered as a secondary source and its energy ${\rm{d}} \mathcal{Q}_s$ can be derived as
\begin{equation}
{\rm{d}} \mathcal{Q}_s=\frac{\mathcal{Q}_t \exp(-k_e \tau) k_s {\rm{d}}v}{2\pi(1-\cos{\beta_t})\tau^2}.
\label{eq:5}
\end{equation} 
Afterward, the scattered energy of ${\rm{d}}v$ collected by the receiver can be expressed as  
\begin{equation}
{\rm{d}}\mathcal{Q}_{r,\rm{sca}}={\rm{d}}\mathcal{Q}_s {\rm{P}}(\cos\vartheta_s)\frac{A_r \cos{\vartheta_v}}{\varepsilon^2}\exp(-k_e \varepsilon).
\label{eq:6}
\end{equation} 
Substituting (\ref{eq:5}) into (\ref{eq:6}) and integrating ${\rm{d}}\mathcal{Q}_{r,\rm{sca}}$ over the whole overlap volume, we can obtain
\begin{subequations}
\begin{equation}
\mathcal{Q}_{r,\rm{sca}}=\int_{\vartheta_{\min}}^{\vartheta_{\max}}\int_{\varpi_{\min}}^{\varpi_{\max}}\int_{\tau_{\min}}^{\tau_{\max}}\mathcal{K}\,\mathcal{G}_{\rm{wei}}\,\rm{d}\tau \rm{d}\varpi \rm{d}\vartheta,
\label{eq:7a}
\end{equation} 
\begin{equation}
\mathcal{K}=\frac{\mathcal{Q}_t\cos{\vartheta_v}{\rm{P}}(\cos{\vartheta_s})\exp[-k_e(\tau+\varepsilon)] A_r k_s |J_3|}{2\pi(1-\cos{\beta_t})\tau^2\varepsilon^2},
\label{eq:7b}
\end{equation}
\end{subequations}
where $\mathcal{G}_{\rm{wei}}$ is the weighting factor to determine whether the scattering point P in the overlap volume is valid or not, and $|J_3|$ can be given by $\tau^2\cos{\varpi}$, whose detailed derivation can be obtained from Appendix~\ref{App:A}.

The unknowns in (\ref{eq:7a}) are the weighting factor and the upper and lower limits of the triple integral, i.e., $\vartheta_{\min}$, $\vartheta_{\max}$, $\varpi_{\min}$, $\varpi_{\max}$, $\tau_{\min}$, and $\tau_{\max}$. To make the entire derivation process easy to understand, the related flow chart is presented in Fig.~\ref{Fig3}.

$\boldsymbol{\delta_t\in[\delta_{t,\rm{low}},\Theta_{t,a}]}$: In this interval, $\vartheta_{\min}$ and $\vartheta_{\max}$ can be expressed as $-\beta_t$ and $\Theta_{t,a}-\vartheta_t$, respectively, and $\varpi_{\min}$ and $\varpi_{\max}$ can be derived as
\begin{equation}
\varpi_{\max}=\tan^{-1}\left(\frac{\sqrt{\tan^2{\beta_t}-\tan^2{\vartheta}}}{\sec{\vartheta}}\right)=-\varpi_{\min}. 
\end{equation}
As for $\tau_{\min}$ and $\tau_{\max}$, they are determined by the intersection situations between the ray TV and the receiver conical surface, and can be derived as
\begin{equation}
[\tau_{\min},\tau_{\max}]=
\begin{cases}
[\tau_0,+\infty), & \tau_0>0,\\ 
[\tau_2,+\infty), & \tau_1<0\,\,\rm{and}\,\,\tau_2>0,\\
[\tau_1,\tau_2], & \tau_1>0,\\
\emptyset, & \rm{otherwise},
\end{cases}
\end{equation}
where the real values $\tau_0$, $\tau_1$, and $\tau_2$ (by default, $\tau_2\ge\tau_1$) can be expressed as
\begin{equation}
\begin{cases}
\tau_0=-\xi_3/\xi_2, &\xi_1=0\,\,\rm{and}\,\,\xi_2\ne0,\\
\tau_{1,2}=\displaystyle{\frac{-\xi_2\pm\sqrt{\Delta}}{2\,\xi_1}}, &\xi_1\ne0\,\,\rm{and}\,\,\Delta\ge0,
\end{cases}
\end{equation} 
and $\Delta=\xi^2_2-4\,\xi_1\xi_3$. Here, $\xi_1$, $\xi_2$, and $\xi_3$ can be derived as
\begin{subequations} 
\begin{equation}
\begin{aligned}
\,\,\xi_1=&-[\cos{\vartheta_r}(\mathcal{A}_1\cos{\alpha_r}+\mathcal{A}_2\sin{\alpha_r})+\mathcal{A}_3\sin{\vartheta_r}]^2\\
                        &+(\mathcal{A}^2_1+\mathcal{A}^2_2+\mathcal{A}^2_3)\cos^2{\beta_r},
\end{aligned}
\end{equation}
\begin{equation}
\begin{aligned}
\xi_2=&\,2\,r[\cos{\vartheta_r}(\mathcal{A}_1\cos{\alpha_r}+\mathcal{A}_2\sin{\alpha_r})+\mathcal{A}_3\sin{\vartheta_r}]\\
                        &\times\cos{\vartheta_r}\sin{\alpha_r}-2\,r \mathcal{A}_2 \cos^2{\beta_r},
\end{aligned}
\end{equation}
\begin{equation}
\xi_3=r^2(\cos^2{\beta_r}-\cos^2{\vartheta_r}\sin^2{\alpha_r}),
\end{equation}
\end{subequations} 
where $\mathcal{A}_1$, $\mathcal{A}_2$, and $\mathcal{A}_3$ are the parametric equation coefficients of $x$, $y$, and $z$ in (\ref{eq:Pzb}), respectively, and $\tau$ is the corresponding parameter. Then, the value of $\mathcal{G}_{\rm{wei}}$ is investigated for different situations. 
\linespread{1.8}
\begin{table}[t]
\centering
\linespread{1.0}
\caption{Relationship Among $\Omega_{t,\rm{min}}$, $\Omega_{t,\rm{max}}$, $\Psi_{t,\rm{min}}$, and $\Psi_{t,\rm{max}}$}
\label{t3}
\begin{tabular}{| c | c |}
\hline
\multirow{2}{*}{\textbf{All Cases}}&$\boldsymbol{\Omega_{t,\rm{min}}}$, $\boldsymbol{\Omega_{t,\rm{max}}}$, $\boldsymbol{\Psi_{t,\rm{min}}}$, \textbf{and} $\boldsymbol{\Psi_{t,\rm{max}}}$\\
{}&\textbf{in descending order}\\
\hline 
{Case 1}&{$\Omega_{t,\rm{max}}>\Omega_{t,\rm{min}}\ge\Psi_{t,\rm{max}}>\Psi_{t,\rm{min}}$}\\
\hline
{Case 2}&{$\Omega_{t,\rm{max}}>\Psi_{t,\rm{max}}>\Omega_{t,\rm{min}}\ge\Psi_{t,\rm{min}}$}\\
\hline
{Case 3}&{$\Omega_{t,\rm{max}}>\Psi_{t,\rm{max}}>\Psi_{t,\rm{min}}\ge\Omega_{t,\rm{min}}$}\\
\hline
{Case 4}&{$\Psi_{t,\rm{max}}\ge\Omega_{t,\rm{max}}>\Omega_{t,\rm{min}}\ge\Psi_{t,\rm{min}}$}\\
\hline
{Case 5}&{$\Psi_{t,\rm{max}}\ge\Omega_{t,\rm{max}}>\Psi_{t,\rm{min}}>\Omega_{t,\rm{min}}$}\\
\hline
{Case 6}&{$\Psi_{t,\rm{max}}>\Psi_{t,\rm{min}}\ge\Omega_{t,\rm{max}}>\Omega_{t,\rm{min}}$}\\
\hline
\end{tabular}
\end{table}
\linespread{1.0} 

First, the intersection circumstances between the transmitter beam and the obstacle must be discussed. Based on geometric relationships, the relationship among $\Omega_{t,\min}$, $\Omega_{t,\max}$, $\Psi_{t,\min}$, and $\Psi_{t,\max}$ are summarized in Table~\ref{t3}, which can be derived as
\begin{subequations}
\begin{equation}
\Omega_{t,\min}=\alpha_t+\varpi_{\min}-\pi/2,
\end{equation} 
\begin{equation}
\Omega_{t,\rm{max}}=\alpha_t+\varpi_{\max}-\pi/2,
\end{equation} 
\begin{equation}
\Psi_{t,\rm{min}}=\min(\Psi_{t,aa'},\Psi_{t,cc'},\Psi_{t,dd'}),
\end{equation} 
\begin{equation}
\Psi_{t,\rm{max}}=\max(\Psi_{t,aa'},\Psi_{t,bb'},\Psi_{t,cc'}).
\end{equation} 
\end{subequations} 
Moreover, the angle $\Psi_{t,nn'}$ between $\boldsymbol{{\rm{TP}}_{t, nn'}}$ and $\boldsymbol{\rm{TK}}$ can be derived as 
\begin{equation}
\Psi_{t,nn'}=\cos^{-1}\left(\frac{y_{nn'}\Xi_{t,c}-z_{t,nn'}\Xi_{t,b}}{\sqrt{\Xi^2_{t,b}+\Xi^2_{t,c}}||\boldsymbol{{\rm{TP}}_{t, nn'}}||}\right),
\end{equation}
where the Z coordinate of ${\rm{P}}_{t,nn'}$ can be expressed as
\begin{equation}
z_{t,nn'}=-(\Xi_{t,a} x_n+\Xi_{t,b} y_n)/\Xi_{t,c},
\end{equation}
and $\Xi_{t,a}$, $\Xi_{t,b}$, and $\Xi_{t,c}$ can be derived as 
\begin{subequations}
\begin{equation}
\Xi_{t,a}=-{\cos{\alpha_t}\sec^2{\vartheta}\sin(2\delta_t)\tan{\varphi_t}},
\end{equation}
\begin{equation}
\Xi_{t,b}=-{\sin{\alpha_t}\sec^2{\vartheta}\sin(2\delta_t)\tan{\varphi_t}},
\end{equation}
\begin{equation}
\Xi_{t,c}={2\sec^2{\vartheta}\cos^2\delta_t\tan{\varphi_t}}.
\end{equation}
\end{subequations}
Meanwhile, $\varphi_t=\tan^{-1}[\sec{\delta_t}(\tan^2\beta_t-\tan^2\vartheta)^{1/2}\cos{\vartheta}]$ and $\vartheta$ $\in(-\beta_t,\beta_t)$.
\linespread{1.8}
\begin{table}[t]
\centering
\linespread{1.0}
\caption{Intersection Orders of the Plane $\mathcal{L}_{\sigma}$ with A, B, C, and D}
\label{t4}
\vspace{2pt}
\begin{tabular}{| c | c |}
\hline
\multirow{2}{*}{\textbf{All Scenarios}}&$\boldsymbol{\Theta_{r,a}}$, $\boldsymbol{\Theta_{r,b}}$, $\boldsymbol{\Theta_{r,c}}$, \textbf{and} $\boldsymbol{\Theta_{r,d}}$\\
{}&\textbf{in descending order}\\
\hline 
{Scenario R1}&{$\Theta_{r,d}>\Theta_{r,a}>\Theta_{r,c}>\Theta_{r,b}$}\\
\hline
{Scenario R2}&{$\Theta_{r,d}>\Theta_{r,c}\ge\Theta_{r,a}>\Theta_{r,b}$}\\
\hline
{Scenario R3}&{$\Theta_{r,d}=\Theta_{r,c}>\Theta_{r,a}=\Theta_{r,b}$}\\
\hline
{Scenario R4}&{$\Theta_{r,a}>\Theta_{r,d}>\Theta_{r,b}>\Theta_{r,c}$}\\
\hline
{Scenario R5}&{$\Theta_{r,a}=\Theta_{r,d}>\Theta_{r,b}=\Theta_{r,c}$}\\
\hline
{Scenario R6}&{$\Theta_{r,c}>\Theta_{r,d}>\Theta_{r,b}>\Theta_{r,a}$}\\
\hline
\end{tabular}
\end{table}
\linespread{1.0}
\linespread{1.8}
\begin{table}[t]
\centering
\linespread{1.0}
\caption{Relationship Among $\Omega_{r,\rm{min}}$, $\Omega_{r,\rm{max}}$, $\Psi_{r,\rm{min}}$, and $\Psi_{r,\rm{max}}$}
\label{t5}
\begin{tabular}{| c | c |}
\hline
\multirow{2}{*}{\textbf{All Conditions}}&$\boldsymbol{\Omega_{r,\rm{min}}}$, $\boldsymbol{\Omega_{r,\rm{max}}}$, $\boldsymbol{\Psi_{r,\rm{min}}}$, \textbf{and} $\boldsymbol{\Psi_{r,\rm{max}}}$\\
{}&\textbf{in descending order}\\
\hline 
{Condition 1}&{$\Omega_{r,\rm{max}}>\Omega_{r,\rm{min}}\ge\Psi_{r,\rm{max}}>\Psi_{r,\rm{min}}$}\\
\hline
{Condition 2}&{$\Omega_{r,\rm{max}}>\Psi_{r,\rm{max}}>\Omega_{r,\rm{min}}\ge\Psi_{r,\rm{min}}$}\\
\hline
{Condition 3}&{$\Omega_{r,\rm{max}}>\Psi_{r,\rm{max}}>\Psi_{r,\rm{min}}>\Omega_{r,\rm{min}}$}\\
\hline
{Condition 4}&{$\Psi_{r,\rm{max}}\ge\Omega_{r,\rm{max}}>\Omega_{r,\rm{min}}\ge\Psi_{r,\rm{min}}$}\\
\hline
{Condition 5}&{$\Psi_{r,\rm{max}}\ge\Omega_{r,\rm{max}}>\Psi_{r,\rm{min}}>\Omega_{r,\rm{min}}$}\\
\hline
{Condition 6}&{$\Psi_{r,\rm{max}}>\Psi_{r,\rm{min}}\ge\Omega_{r,\rm{max}}>\Omega_{r,\rm{min}}$}\\
\hline
\end{tabular}
\end{table}
\linespread{1.0} 

Second, whether the scattered photon arrives at the receiver successfully must be discussed. As $\mathcal{L}_{\sigma}$ rotates clockwise from $\mathcal{L}_{-\vartheta_r}$ to $\mathcal{L}_{\pi/2-\vartheta_r}$, the possible intersection orders of $\mathcal{L}_{\sigma}$ with A, B, C, and D are summarized in Table~\ref{t4}, where $\Theta_{r,a}$, $\Theta_{r,b}$, $\Theta_{r,c}$, and $\Theta_{r,d}$ can be obtained by
\begin{equation}
\Theta_{r,n}=\tan^{-1}\left(\frac{\kappa\sqrt{\cot^2{\alpha_r}+1}}{|x_n\cot{\alpha_r}+y_n-r|}\right).
\end{equation} 
Given that Scenario R1 is the most common circumstance, we take it as an example to carry out the detailed derivation. The intersection circumstances between the receiver FoV and the obstacle are presented in Table~\ref{t5}, where $\Omega_{r,\rm{min}}=-\alpha_r-\pi/2-\mathcal{C}$, $\Omega_{r,\rm{max}}=-\alpha_r-\pi/2+\mathcal{C}$, and $\mathcal{C}$ can be derived as 
\begin{equation}
\mathcal{C}=\tan^{-1}\left(\cos{\sigma}\sqrt{\tan^2{\beta_r}-\tan^2{\sigma}}\right).
\end{equation} 
The derivation process for $\Psi_{r,\rm{min}}$ and $\Psi_{r,\rm{max}}$ can be obtained from Appendix~\ref{App:B}, where $\sigma\in(-\beta_r,\beta_r)$.

Since $\vartheta$ and $\varpi$ are known, the intersection situations of TV with the receiver conical surface can be easily determined. If $[\tau_{\min},\tau_{\max}]=\emptyset$, $\mathcal{G}_{\rm{wei}}=0$, which means that the emitted ray TV has no energy contribution to $\mathcal{Q}_{r,\rm{sca}}$. If $[\tau_{\min},\tau_{\max}]\neq\emptyset$, the value of $\mathcal{G}_{\rm{wei}}$ is closely related to the value of $\delta_r$, where the effective scattering point is located on the corresponding plane $\mathcal{L}_{\sigma}$. If $\delta_r\ge\Theta_{r,d}$, $\mathcal{G}_{\rm{wei}}=1$, while if $\delta_r<\Theta_{r,d}$, the~value of $\mathcal{G}_{\rm{wei}}$ will be investigated from \textbf{Case 1} to \textbf{Case 5}. For \textbf{Case 6} and {\em{Condition 6}}, the value of $\mathcal{G}_{\rm{wei}}$ is always equal to $1$, since the propagation links of the UV signal are not blocked by the obstacle. For clarity, we only provide the circumstances where the value of $\mathcal{G}_{\rm{wei}}$ is equal to $1$ in the following derivation, and $\mathcal{G}_{\rm{wei}}=0$ in other circumstances.

\textbf{Case 1:} $\Omega_{t,\rm{max}}>\Omega_{t,\rm{min}}\ge\Psi_{t,\rm{max}}>\Psi_{t,\rm{min}}$

In this case, $\mathcal{G}_{\rm{wei}}$ is always equal to $1$ for {\em{Condition 1}}. For {\em{Condition 2}}, $\Psi_{r,\rm{esp}}\in(\Psi_{r,\rm{max}},\Omega_{r,\rm{max}}]$, where $\Psi_{r,\rm{esp}}$ denotes the angle between $\boldsymbol{\rm{RS}}$ and $\boldsymbol{\rm{RP}}$, while for {\em{Condition 3}}, $\Psi_{r,\rm{esp}}$ $\in[\Omega_{r,\rm{min}},\Psi_{r,\rm{min}})\cup(\Psi_{r,\rm{max}},\Omega_{r,\rm{max}}]$. Further, $\Psi_{r,\rm{esp}}$ belongs to the interval $[\Omega_{r,\rm{min}},\Psi_{r,\rm{min}})$ for {\em{Condition 5}}.  

\textbf{Case 2:} $\Omega_{t,\rm{max}}>\Psi_{t,\rm{max}}>\Omega_{t,\rm{min}}\ge\Psi_{t,\rm{min}}$

In this case, the situations where $\mathcal{G}_{\rm{wei}}$ is always equal to $1$ are presented first. For {\em{Condition 1}}, $\Psi_{t,\rm{esp}}\in(\Psi_{t,\rm{max}},\Omega_{t,\rm{max}}]$, where $\Psi_{t,\rm{esp}}$ is the angle between $\boldsymbol{\rm{TK}}$ and $\boldsymbol{\rm{TP}}$, while for {\em{Conditions 2}} and {\em{3}}, $\Psi_{t,\rm{esp}}$ belongs to $(\Psi_{t,\rm{max}},\Omega_{t,\rm{max}}]$ and $\Psi_{r,\rm{esp}}$ belongs to $(\Psi_{r,\rm{max}},\Omega_{r,\rm{max}}]$ simultaneously. Besides, the circumstance where $\Psi_{r,\rm{esp}}\in[\Omega_{r,\rm{min}},\Psi_{r,\rm{min}})$ needs to be incorporated for {\em{Condition 3}}, and this interval also applies to {\em{Condition 5}}.
\linespread{1.7}
\begin{table}[t]
\centering
\linespread{1.0}
\caption{Situations I-1 to I-9}
\label{t6}
\begin{tabular}{ c  c  c  c  c  }
\hline
\hline
\textbf{All Situations}&$\boldsymbol{\Psi_{t,\max}}$&$\boldsymbol{\Psi_{t,\min}}$&$\boldsymbol{\Psi_{r,\max}}$&$\boldsymbol{\Psi_{r,\min}}$\\
\hline
\hline
Situation I-1 & $\Psi_{t,bb'}$ & $\Psi_{t,dd'}$ & $\Psi_{r,aa'}$ & $\Psi_{r,cc'}$\\      

Situation I-2 & $\Psi_{t,cc'}$ & $\Psi_{t,dd'}$ & $\Psi_{r,aa'}$ & $\Psi_{r,cc'}$\\      

Situation I-3 & $\Psi_{t,bb'}$ & $\Psi_{t,dd'}$ & $\Psi_{r,dd'}$ & $\Psi_{r,cc'}$\\      

Situation I-4 & $\Psi_{t,cc'}$ & $\Psi_{t,aa'}$ & $\Psi_{r,bb'}$ & $\Psi_{r,dd'}$\\      

Situation I-5 & $\Psi_{t,cc'}$ & $\Psi_{t,aa'}$ & $\Psi_{r,aa'}$ & $\Psi_{r,dd'}$\\      

Situation I-6 & $\Psi_{t,cc'}$ & $\Psi_{t,aa'}$ & $\Psi_{r,aa'}$ & $\Psi_{r,cc'}$\\      

Situation I-7 & $\Psi_{t,bb'}$ & $\Psi_{t,cc'}$ & $\Psi_{r,dd'}$ & $\Psi_{r,bb'}$\\      

Situation I-8 & $\Psi_{t,aa'}$ & $\Psi_{t,cc'}$ & $\Psi_{r,dd'}$ & $\Psi_{r,bb'}$\\      

Situation I-9 & $\Psi_{t,bb'}$ & $\Psi_{t,dd'}$ & $\Psi_{r,dd'}$ & $\Psi_{r,bb'}$\\      
\hline
\hline
\end{tabular}
\end{table}
\linespread{1.0}

Second, all possible circumstances where $\mathcal{G}_{\rm{wei}}$ is equal to~$1$ are summarized in Table~\ref{t6}, and meanwhile, three constraints imposed on $||\boldsymbol{\rm{TP}}||$ for different values of $\alpha$ are summarized~as
\begin{subequations}
\begin{align}
\mathbb{T}1: &<\frac{||\boldsymbol{{\rm{TP}}_{t,dd'}}||\sin{\psi_{t,dd'}}}{\sin(\Psi_{t,\rm{esp}}+\psi_{t,dd'}-\Psi_{t,dd'})}, \label{eq:TPCD}\\ 
\mathbb{T}2: &<\frac{||\boldsymbol{{\rm{TP}}_{t,aa'}}||\sin{\psi_{t,aa'}}}{\sin(\Psi_{t,\rm{esp}}+\psi_{t,aa'}-\Psi_{t,aa'})}, \label{eq:TPAD} \\ 
\mathbb{T}3: &<\frac{||\boldsymbol{{\rm{TP}}_{t,cc'}}||\sin{\psi_{t,cc'}}}{\sin(\Psi_{t,\rm{esp}}+\psi_{t,cc'}-\Psi_{t,cc'})}, \label{eq:TPBC} 
\end{align}
\end{subequations}
where $\psi_{t,dd'}$, $\psi_{t,aa'}$, and $\psi_{t,cc'}$ represent the angles between $\boldsymbol{{\rm{TP}}_{t,dd'}}$ and $\boldsymbol{{\rm{P}}_{t,cc'}{\rm{P}}_{t,dd'}}$, $\boldsymbol{{\rm{TP}}_{t, a a'}}$ and $\boldsymbol{{\rm{P}}_{t, d d'}{\rm{P}}_{t, a a'}}$, and $\boldsymbol{{\rm{TP}}_{t,cc'}}$ and $\boldsymbol{{\rm{P}}_{t,bb'}{\rm{P}}_{t,cc'}}$, respectively. Then, the constraints imposed on $||\boldsymbol{\rm{RP}}||$ for different $\alpha$ and $\delta_r$ values are presented as
\begin{subequations}
\begin{align}
\mathbb{R}1: &<\frac{||\boldsymbol{{\rm{RP}}_{r,cc'}}||\sin{\psi_{r,cc'}}}{\sin(\Psi_{r,\rm{esp}}+\psi_{r,cc'}-\Psi_{r,cc'})}, \label{eq:RPCD}\\ 
\mathbb{R}2: &<\frac{||\boldsymbol{{\rm{RP}}_{r,cd}}||\sin{\psi_{r,cd}}}{\sin(\Psi_{r,\rm{esp}}+\psi_{r,cd}-\Psi_{r,cd})}, \label{eq:RPC_D}\\ 
\mathbb{R}3: &<\frac{||\boldsymbol{{\rm{RP}}_{r,dd'}}||\sin{\psi_{r,dd'}}}{\sin(\Psi_{r,\rm{esp}}+\psi_{r,dd'}-\Psi_{r,dd'})}, \label{eq:RPAD}\\
\mathbb{R}4: &<\frac{||\boldsymbol{{\rm{RP}}_{r,dd'}}||\sin{\psi_{r,ad}}}{\sin(\Psi_{r,\rm{esp}}+\psi_{r,ad}-\Psi_{r,dd'})}, \label{eq:RPA_D}\\
\mathbb{R}5: &<\frac{||\boldsymbol{{\rm{RP}}_{r,bb'}}||\sin{\psi_{r,bb'}}}{\sin(\Psi_{r,\rm{esp}}+\psi_{r,bb'}-\Psi_{r,bb'})}, \label{eq:RPBC}\\
\mathbb{R}6: &<\frac{||\boldsymbol{{\rm{RP}}_{r,bc}}||\sin{\psi_{r,bc}}}{\sin(\Psi_{r,\rm{esp}}+\psi_{r,bc}-\Psi_{r,bc})}, \label{eq:RPB_C}
\end{align}
\end{subequations}
where $\psi_{r,cc'}$, $\psi_{r,cd}$, $\psi_{r,dd'}$, $\psi_{r,ad}$, $\psi_{r,bb'}$, and $\psi_{r,bc}$ represent the angles between $\boldsymbol{{\rm{P}}_{r,dd'}{\rm{P}}_{r,cc'}}$ and $\boldsymbol{{\rm{RP}}_{r,cc'}}$, $\boldsymbol{{\rm{P}}_{r, d d'}{\rm{P}}_{r, c d}}$ and $\boldsymbol{{\rm{RP}}_{r, c d}}$, $\boldsymbol{{\rm{P}}_{r, a a'}{\rm{P}}_{r, d d'}}$ and $\boldsymbol{{\rm{RP}}_{r, d d'}}$, $\boldsymbol{{\rm{P}}_{r, a d}{\rm{P}}_{r, d d'}}$ and $\boldsymbol{{\rm{RP}}_{r, d d'}}$, $\boldsymbol{{\rm{P}}_{r,cc'}{\rm{P}}_{r,bb'}}$ and $\boldsymbol{{\rm{RP}}_{r,bb'}}$, and $\boldsymbol{{\rm{P}}_{r,cc'}{\rm{P}}_{r,bc}}$ and $\boldsymbol{{\rm{RP}}_{r,bc}}$, respectively. 

1) In situations I-1, I-2, and I-3, $\Psi_{t,\rm{esp}}\in[\Omega_{t,\min},\Psi_{t,cc'}]^{\mathbb{T}1}$ and $\Psi_{r,\rm{esp}}\in[\Omega_{r,\min},\Psi_{r,dd'}]^{\mathbb{R}1}$ for {\em{Condition 2}}, where the superscripts ``$\mathbb{T}1$'' and ``$\mathbb{R}1$'' represent the constraint number of $||\boldsymbol{\rm{TP}}||$ and $||\boldsymbol{\rm{RP}}||$ that match the relevant intervals, while for {\em{Conditions 3}}, {\em{4}}, and {\em{5}}, related $\Psi_{r,\rm{esp}}$ intervals are changed to $[\Psi_{r,\min},\Psi_{r,dd'}]^{\mathbb{R}1}$, $[\Omega_{r,\min},\min(\Omega_{r,\max},\Psi_{r,dd'})]^{\mathbb{R}1}$, and $[\Psi_{r,\min},\min(\Omega_{r,\max},\Psi_{r,dd'})]^{\mathbb{R}1}$, respectively. If $\delta_r>\Theta_{r,c}$, the superscript ``$\mathbb{R}1$'' needs to be changed to ``$\mathbb{R}2$''. During the whole analysis, the values of $\Psi_{t,\min}$, $\Psi_{t,\max}$, $\Psi_{r,\min}$, and $\Psi_{r,\max}$ are presented with reference to the case where $\delta_r\le\Theta_{r,b}$. Note that in situations I-1 and I-3, $\Omega_{t,\min}$ is assumed to be less than $\Psi_{t,cc'}$.

2) In situations I-4 and I-5, $\Psi_{t,\rm{esp}}\in[\Omega_{t,\min},\Psi_{t,dd'}]^{\mathbb{T}2}$. For {\em{Conditions 2}}, {\em{3}}, {\em{4}}, and {\em{5}}, $\Psi_{r,\rm{esp}}\in[\Omega_{r,\min},\Psi_{r,aa'}]^{\mathbb{R}3}$, $[\Psi_{r,\min},$ $\Psi_{r,aa'}]^{\mathbb{R}3}$, $[\Omega_{r,\min},\min(\Omega_{r,\max},\Psi_{r,aa'})]^{\mathbb{R}3}$, and $[\Psi_{r,\min},\min$ $(\Omega_{r,\max},\Psi_{r,aa'})]^{\mathbb{R}3}$, respectively, where the superscript ``$\mathbb{R}3$'' is changed to ``$\mathbb{R}4$'' when $\delta_r>\Theta_{r,a}$. Note that in situations I-4 and I-5, $\Omega_{t,\min}$ is assumed to be less than $\Psi_{t,dd'}$.

3) In the situation I-6, $\Psi_{t,\rm{esp}}\in[\Omega_{t,\min},\Psi_{t,dd'}]^{\mathbb{T}2}\cup(\Psi_{t,dd'},$ $\Psi_{t,\max}]^{\mathbb{T}1}$, and for {\em{Condition 2}}, $\Psi_{r,\rm{esp}}\in[\Omega_{r,\min},\Psi_{r,\max}]^{\mathbb{R}3}$ if $\Omega_{r,\min}\ge\Psi_{r,dd'}$, and $\Psi_{r,\rm{esp}}\in[\Omega_{r,\min},\Psi_{r,dd'}]^{\mathbb{R}1}\cup(\Psi_{r,dd'},$ $\Psi_{r,\max}]^{\mathbb{R}3}$ if $\Omega_{r,\min}<\Psi_{r,dd'}$. As for {\em{Condition 3}}, $\Psi_{r,\rm{esp}}\in$ $[\Psi_{r,\min},\Psi_{r,dd'}]^{\mathbb{R}1}\cup(\Psi_{r,dd'},\Psi_{r,\max}]^{\mathbb{R}3}$, while for {\em{Condition~4}}, $\Psi_{r,\rm{esp}}\in[\Omega_{r,\min},\Omega_{r,\max}]^{\mathbb{R}1}$, $[\Omega_{r,\min},\Omega_{r,\max}]^{\mathbb{R}3}$, and $[\Omega_{r,\min},$ $\Psi_{r,dd'}]^{\mathbb{R}1}\cup(\Psi_{r,dd'},\Omega_{r,\max}]^{\mathbb{R}3}$ for $\Omega_{r,\max}\le\Psi_{r,dd'}$, $\Omega_{r,\min}\ge\Psi_{r,dd'}$, and $\Omega_{r,\min}<\Psi_{r,dd'}<\Omega_{r,\max}$, respectively. Regarding {\em{Condition~5}}, $\Psi_{r,\rm{esp}}\in[\Psi_{r,\min},\Psi_{r,dd'}]^{\mathbb{R}1}\cup(\Psi_{r,dd'},$ $\Omega_{r,\max}]^{\mathbb{R}3}$ for $\Omega_{r,\max}>\Psi_{r,dd'}$ and $[\Psi_{r,\min},\Omega_{r,\max}]^{\mathbb{R}1}$ otherwise. If $\delta_r>\Theta_{r,c}$ (or $\Theta_{r,a}$), the superscript ``$\mathbb{R}1$'' (or ``$\mathbb{R}3$'') is changed to ``$\mathbb{R}2$'' (or ``$\mathbb{R}4$''). In this situation, $\Omega_{t,\min}$ is assumed to be less than $\Psi_{t,dd'}$, and when $\Omega_{t,\min}\ge\Psi_{t,dd'}$, the derivation of $\mathcal{G}_{\rm{wei}}$ is in accordance with that of I-1, I-2, and I-3.

4) In situations I-7 and I-8, $\Psi_{t,\rm{esp}}\in[\Omega_{t,\min},\Psi_{t,bb'}]^{\mathbb{T}3}$ and $\Psi_{r,\rm{esp}}\in[\Omega_{r,\min},\Psi_{r,cc'}]^{\mathbb{R}5}$ for {\em{Condition 2}}, while for {\em{Conditions 3}}, {\em{4}}, and {\em{5}}, $\Psi_{r,\rm{esp}}$ belongs to $[\Psi_{r,\min},\Psi_{r,cc'}]^{\mathbb{R}5}$, $[\Omega_{r,\min},\min(\Omega_{r,\max},\Psi_{r,cc'})]^{\mathbb{R}5}$, and $[\Psi_{r,\min},\min(\Omega_{r,\max},$ $\Psi_{r,cc'})]^{\mathbb{R}5}$, respectively. If $\Theta_{r,b},<\delta_r<\Theta_{r,c}$, the superscript ``$\mathbb{R}5$'' is changed to ``$\mathbb{R}6$'' for relevant intervals. In I-8, $\Omega_{t,\min}$ is assumed to be less than $\Psi_{t,bb'}$. 

5) In the situation I-9, $\Psi_{t,\rm{esp}}\in[\Omega_{t,\min},\Psi_{t,cc'}]^{\mathbb{T}1}\cup(\Psi_{t,cc'},$ $\Psi_{t,\max}]^{\mathbb{T}3}$, and for {\em{Condition 2}}, $\Psi_{r,\rm{esp}}\in[\Omega_{r,\min},\Psi_{r,\max}]^{\mathbb{R}1}$ if $\Omega_{r,\min}\ge\Psi_{r,cc'}$, and $\Psi_{r,\rm{esp}}\in[\Omega_{r,\min},\Psi_{r,cc'}]^{\mathbb{R}5}\cup(\Psi_{r,cc'},$ $\Psi_{r,\max}]^{\mathbb{R}1}$ if $\Omega_{r,\min}<\Psi_{r,cc'}$. As for {\em{Condition 3}}, $\Psi_{r,\rm{esp}}\in$ $[\Psi_{r,\min},\Psi_{r,cc'}]^{\mathbb{R}5}\cup(\Psi_{r,cc'},\Psi_{r,\max}]^{\mathbb{R}1}$, while for {\em{Condition~4}}, $\Psi_{r,\rm{esp}}\in[\Omega_{r,\min},\Omega_{r,\max}]^{\mathbb{R}5}$, $[\Omega_{r,\min},\Omega_{r,\max}]^{\mathbb{R}1}$, and $[\Omega_{r,\min},$ $\Psi_{r,cc'}]^{\mathbb{R}5}\cup(\Psi_{r,cc'},\Omega_{r,\max}]^{\mathbb{R}1}$ for $\Omega_{r,\max}\le\Psi_{r,cc'}$, $\Omega_{r,\min}\ge\Psi_{r,cc'}$, and $\Omega_{r,\min}<\Psi_{r,cc'}<\Omega_{r,\max}$, respectively. Regarding {\em{Condition~5}}, $\Psi_{r,\rm{esp}}\in[\Psi_{r,\min},\Psi_{r,cc'}]^{\mathbb{R}5}\cup(\Psi_{r,cc'},$ $\Omega_{r,\max}]^{\mathbb{R}1}$ for $\Omega_{r,\max}>\Psi_{r,cc'}$ and $[\Psi_{r,\min},\Omega_{r,\max}]^{\mathbb{R}5}$ otherwise. If $\delta_r>\Theta_{r,b}$ ($\Theta_{r,c}$), the superscript ``$\mathbb{R}5$'' (``$\mathbb{R}1$'') is changed to ``$\mathbb{R}6$'' (``$\mathbb{R}2$''). In this situation, $\Omega_{t,\min}$ is supposed to be less than $\Psi_{t,cc'}$, and when $\Omega_{t,\min}\ge\Psi_{t,cc'}$, the derivation of $\mathcal{G}_{\rm{wei}}$ is in accordance with that of I-7 and I-8. Note that throughout the whole modeling process, the $\Psi_{r,\rm{esp}}$ intervals must be updated when their superscripts are changed.  

\textbf{Case 3}: $\Omega_{t,\rm{max}}>\Psi_{t,\rm{max}}>\Psi_{t,\rm{min}}\ge\Omega_{t,\rm{min}}$

In this case, the circumstances where $\mathcal{G}_{\rm{wei}}$ is always equal to $1$ are provided first. For {\em{Condition~1}}, $\Psi_{t,\rm{esp}}\in[\Omega_{t,\min},\Psi_{t,\min})$ $\cup\,(\Psi_{t,\max},\Omega_{t,\max}]$, while for {\em{Conditions~2}}, {\em{3}}, {\em{4}}, and {\em{5}}, $\Psi_{t,\rm{esp}}$ $\in[\Omega_{t,\min},\Psi_{t,\min})$. Additionally, the situation where $\Psi_{t,\rm{esp}}\in$ $(\Psi_{t,\max},\Omega_{t,\max}]$ and $\Psi_{r,\rm{esp}}\in(\Psi_{r,\max},\Omega_{r,\max}]$ for {\em{Condition 2}} needs to be considered, while for {\em{Condition 5}}, the situation where $\Psi_{t,\rm{esp}}\in[\Psi_{t,\min},\Omega_{t,\max}]$ and $\Psi_{r,\rm{esp}}\in[\Omega_{r,\min},\Psi_{r,\min})$ must be incorporated, and for {\em{Condition 3}}, the following parts: i) $\Psi_{r,\rm{esp}}\in[\Omega_{r,\min},\Psi_{r,\min})$ and $\Psi_{t,\rm{esp}}\in[\Psi_{t,\min},\Psi_{t,\max}]$ and ii) $\Psi_{r,\rm{esp}}\in[\Omega_{r,\min},\Psi_{r,\min})\cup(\Psi_{r,\max},\Omega_{r,\max}]$ and $\Psi_{t,\rm{esp}}\in$ $(\Psi_{t,\max},\Omega_{t,\max}]$ must be taken into account. For the possible circumstances where $\mathcal{G}_{\rm{wei}}=1$, they are consistent with those of \textbf{Case 2}, where the item ``$\Omega_{t,\min}$'' of the corresponding $\Psi_{t,\rm{esp}}$ intervals requires to be changed to ``$\Psi_{t,\min}$''. Meanwhile, the constraints imposed on $\Omega_{t,\min}$ become invalid.

\textbf{Case 4}: $\Psi_{t,\rm{max}}\ge\Omega_{t,\rm{max}}>\Omega_{t,\rm{min}}\ge\Psi_{t,\rm{min}}$

When $\Psi_{r,\rm{esp}}\in[\Omega_{r,\min},\Psi_{r,\min})$ in this case, $\mathcal{G}_{\rm{wei}}$ is always equal to $1$ for {\em{Conditions~3}} and {\em{5}}. Following that, the possible circumstances where $\mathcal{G}_{\rm{wei}}=1$ are supplemented. In situations I-1, I-2, and I-3, $\Psi_{t,\rm{esp}}\in[\Omega_{t,\min},\min(\Omega_{t,\max},\Psi_{t,cc'})]^{\mathbb{T}1}$, and in I-4 and I-5, $\Psi_{t,\rm{esp}}\in[\Omega_{t,\min},\min(\Omega_{t,\max},\Psi_{t,dd'})]^{\mathbb{T}2}$, while in I-6, $\Psi_{t,\rm{esp}}\in[\Omega_{t,\min},\Omega_{t,\max}]^{\mathbb{T}2}$, $[\Omega_{t,\min},\Omega_{t,\max}]^{\mathbb{T}1}$, and $[\Omega_{t,\min},\Psi_{t,dd'}]^{\mathbb{T}2}\cup(\Psi_{t,dd'},\Omega_{t,\max}]^{\mathbb{T}1}$ for $\Omega_{t,\max}\le\Psi_{t,dd'}$, $\Omega_{t,\min}\ge\Psi_{t,dd'}$, and $\Omega_{t,\min}<\Psi_{t,dd'}<\Omega_{t,\max}$, respectively. In I-7 and I-8, $\Psi_{t,\rm{esp}}\in[\Omega_{t,\min},\min(\Omega_{t,\max},\Psi_{t,bb'})]^{\mathbb{T}3}$, while I-9, $\Psi_{t,\rm{esp}}\in[\Omega_{t,\min},\Omega_{t,\max}]^{\mathbb{T}1}$, $[\Omega_{t,\min},\Omega_{t,\max}]^{\mathbb{T}3}$, and $[\Omega_{t,\min},\Psi_{t,cc'}]^{\mathbb{T}1}\cup(\Psi_{t,cc'},\Omega_{t,\max}]^{\mathbb{T}3}$ for $\Omega_{t,\max}\le\Psi_{t,cc'}$, $\Omega_{t,\min}\ge\Psi_{t,cc'}$, and $\Omega_{t,\min}<\Psi_{t,cc'}<\Omega_{t,\max}$, respectively. As for the corresponding $\Psi_{r,\rm{esp}}$ intervals, they are consistent with those of \textbf{Case 2}, where the constraints imposed on $\Omega_{t,\min}$ become invalid.  

\textbf{Case 5}: $\Psi_{t,\rm{max}}\ge\Omega_{t,\rm{max}}>\Psi_{t,\rm{min}}>\Omega_{t,\rm{min}}$ 
 
In this case, the situations where $\mathcal{G}_{\rm{wei}}$ is always equal to $1$ are summarized first. When $\Psi_{t,\rm{esp}}\in[\Omega_{t,\min},\Psi_{t,\min})$, $\mathcal{G}_{\rm{wei}}=1$ for {\em{Conditions~1}} to {\em{5}}, and when $\Psi_{t,\rm{esp}}\in[\Psi_{t,\min},\Omega_{t,\max}]$ and $\Psi_{r,\rm{esp}}\in[\Omega_{r,\min},\Psi_{r,\min})$, $\mathcal{G}_{\rm{wei}}=1$ for {\em{Conditions~3}} and {\em{5}}. Second, the possible circumstances where $\mathcal{G}_{\rm{wei}}$ is equal to $1$ are supplemented. In I-1, I-2, and I-3, $\Psi_{t,\rm{esp}}$ belongs to the interval $[\Psi_{t,\min},\min(\Omega_{t,\max},\Psi_{t,cc'})]^{\mathbb{T}1}$, while in I-4 and I-5, $\Psi_{t,\rm{esp}}$ belongs to the interval $[\Psi_{t,\min},\min(\Omega_{t,\max},\Psi_{t,dd'})]^{\mathbb{T}2}$, and in I-6, $\Psi_{t,\rm{esp}}\in[\Psi_{t,\min},\Psi_{t,dd'}]^{\mathbb{T}2}\cup(\Psi_{t,dd'},\Omega_{t,\max}]^{\mathbb{T}1}$ for $\Omega_{t,\max}>\Psi_{t,dd'}$ and $\Psi_{t,\rm{esp}}\in[\Psi_{t,\min},\Omega_{t,\max}]^{\mathbb{T}2}$ for $\Omega_{t,\max}\le\Psi_{t,dd'}$. In regard to I-7 and I-8, $\Psi_{t,\rm{esp}}$ belongs to the interval $[\Psi_{t,\min},\min(\Omega_{t,\max},\Psi_{t,bb'})]^{\mathbb{T}3}$, and in I-9, $\Psi_{t,\rm{esp}}$ belongs to the intervals $[\Psi_{t,\min},\Psi_{t,cc'}]^{\mathbb{T}1}\cup(\Psi_{t,cc'},\Omega_{t,\max}]^{\mathbb{T}3}$ for $\Omega_{t,\max}$ $>\Psi_{t,cc'}$ and the interval $[\Psi_{t,\min},\Omega_{t,\max}]^{\mathbb{T}1}$ for $\Omega_{t,\max}\le\Psi_{t,cc'}$. With respect to the relevant $\Psi_{r,\rm{esp}}$ intervals, they are the same as those of \textbf{Case 2}, where the limitations imposed on $\Omega_{t,\min}$ become invalid.

$\boldsymbol{\delta_t\in(\Theta_{t,a},\Theta_{t,c})}$: The derivation process for the unspecified quantities in (\ref{eq:7a}) is in accordance with that of the interval $\boldsymbol{(\delta_{t,\rm{low}},\Theta_{t,a}]}$ and thus is not repeated, where the derivations of $\Psi_{t,\min}$ and $\Psi_{t,\max}$ are consistent with those of $\Psi_{r,\min}$ and $\Psi_{r,\max}$, and the equation of the plane $\mathcal{F}_{\vartheta}$ can be expressed as $\Xi_{t,a}x+\Xi_{t,b}y+\Xi_{t,c}z=0$.

$\boldsymbol{\delta_t\in[\Theta_{t,c},\delta_{t,\rm{high}}]}$: In this situation, only the intersection circumstances between the receiver FoV and the obstacle need to be incorporated, because the transmitter beam bypasses the obstacle completely overhead. Note that in the above analysis, we did not consider the contribution of the beam ray $\mathbb{L}^{t}_{-}$ and FoV rays $\mathbb{L}^{r}_\pm$ to the received energy $\mathcal{Q}_{r,\rm{sca}}$, because in these scenarios, $\Omega_{t,\max}=\Omega_{t,\min}$ or $\Omega_{r,\max}=\Omega_{r,\min}$, where $\mathbb{L}^{t}_{-}$ is the intersection ray between the plane $\mathcal{F}_{-\beta_t}$ and the transmitter beam, and $\mathbb{L}^{r}_\pm$ represent the intersection rays between planes $\mathcal{L}_{\pm\beta_r}$ and the receiver FoV. Compared to the situations where $\delta_t\in(\delta_{t,\rm{low}},\Theta_{t,c})$ and $\delta_r\in(\delta_{r,\rm{low}},\Theta_{r,d})$, the situations where $\delta_t=\delta_{t,\rm{low}}$ or $\delta_r=\delta_{r,\rm{low}}$($\delta_{r,\rm{high}}$) are the simple cases, whose energy contribution to $\mathcal{Q}_{r,\rm{sca}}$ can be obtained easily referring to the above modeling. For tractable analysis, the equations of $\mathbb{L}^{t}_{-}$ and $\mathbb{L}^{r}_\pm$ are provided as 
\begin{align}
\mathbb{L}^{t}_{-}:& 
\begin{cases}
x=\sec{\beta_t}\cos\delta_{t,\rm{low}}\cos{\alpha_t}\,\omega,\\
y=\sec{\beta_t}\cos\delta_{t,\rm{low}}\sin{\alpha_t}\,\omega,\\
z=\sec{\beta_t}\sin\delta_{t,\rm{low}}\,\omega,\\
\end{cases}\\
\mathbb{L}^{r}_\pm: &
\begin{cases}
x=\sec{\beta_r}\cos\delta_{r,m\pm}\cos{\alpha_r}\,\omega,\\
y=\sec{\beta_r}\cos\delta_{r,m\pm}\sin{\alpha_r}\,\omega+r,\\
z=\sec{\beta_r}\sin\delta_{r,m\pm}\,\omega,
\end{cases}
\end{align}
where $\omega$ is the parameter, and $\delta_{r,m-}$ and $\delta_{r,m+}$ denote $\delta_{r,\rm{low}}$ and $\delta_{r,\rm{high}}$, respectively. The equation of $\mathcal{L}_{\sigma}$ can be expressed as
\begin{equation}
\Xi_{r,a}x+\Xi_{r,b}(y-r)+\Xi_{r,c}z=0,
\end{equation}
and the equation of the receiver conical surface can be given~by
\begin{equation} 
\cos^2{\beta_r}=\frac{(\mathcal{N}_{r,x} x+\mathcal{N}_{r,y} y+\mathcal{N}_{r,z} z-\mathcal{N}_{r,y} r)^2}{x^2+(y-r)^2+z^2},     
\end{equation}
where $\mathcal{N}_{r,x}=\cos{\vartheta_r}\cos{\alpha_r}$, $\mathcal{N}_{r,y}=\cos{\vartheta_r}\sin{\alpha_r}$, and $\mathcal{N}_{r,z}=\sin{\vartheta_r}$. Since the channel modeling process of Scenarios T2 to T6 is similar to that of Scenario T1, we do not repeat them in this paper. 

\section{Modeling of UV NLoS Reflection Channel Incorporating An Obstacle}

From Fig.~\ref{Fig2} it can be discovered that there are three potential reflection surfaces during the rotation of the obstacle, namely surface $\rm{CDD'C'}$, surface $\rm{DAA'D'}$, and surface $\rm{BCC'B'}$. Next, we first determine the effective reflection regions for the three surfaces, namely, $\mathbb{U}_{\rm{cdd'c'}}$, $\mathbb{U}_{\rm{daa'd'}}$, and $\mathbb{U}_{\rm{bcc'b'}}$.

According to Fig.~\ref{Fig2}, the effective reflection regions $\mathbb{U}_{\rm{cdd'c'}}$, $\mathbb{U}_{\rm{daa'd'}}$, and $\mathbb{U}_{\rm{bcc'b'}}$ can be given by
\begin{align}
\mathbb{U}_{\rm{cdd'c'}}: &
\begin{cases}
y_c \le y \le y_d,\\
\beta_{t,\rm{esp}} \le \beta_t,\\
x=
\begin{cases}
k_{cd}+x_d,\,\alpha \ne 0,\\
x_c,\,\alpha=0,
\end{cases}\\
\beta_{r,\rm{esp}} \le \beta_r,\\
z \le \kappa,
\end{cases}\\
\mathbb{U}_{\rm{daa'd'}}: &
\begin{cases}
y_d \le y \le y_a,\\
\beta_{t,\rm{esp}} \le \beta_t,\\
x=k_{ad}+x_d,\,\alpha < 0,\\
\beta_{r,\rm{esp}} \le \beta_r,\\
z \le \kappa,
\end{cases}\\
\mathbb{U}_{\rm{bcc'b'}}: &
\begin{cases}
y_b \le y \le y_c,\\
\beta_{t,\rm{esp}} \le \beta_t,\\
x=k_{cb}+x_b,\,\alpha > 0,\\
\beta_{r,\rm{esp}} \le \beta_r,\\
z \le \kappa,
\end{cases}
\end{align}
where $\beta_{t,\rm{esp}}$ and $\beta_{r,\rm{esp}}$ can be expressed as
\begin{subequations}
\begin{equation} 
\cos\beta_{t,\rm{esp}}=(\mathcal{N}_{t,x}\,x+\mathcal{N}_{t,y}\,y+\mathcal{N}_{t,z}\,z)/\tau,  
\end{equation} 
\begin{equation} 
\cos\beta_{r,\rm{esp}}=(\mathcal{N}_{r,x}\,x+\mathcal{N}_{r,y}\,(y-r)+\mathcal{N}_{r,z}\,z)/\varepsilon,   
\end{equation} 
and
\begin{equation} 
k_{cd}=\frac{(x_c-x_d)(y-y_d)}{(y_c-y_d)}, 
\end{equation} 
\begin{equation} 
k_{ad}=\frac{(x_a-x_d)(y-y_d)}{(y_a-y_d)},
\end{equation} 
\begin{equation} 
k_{cb}=\frac{(x_c-x_b)(y-y_b)}{(y_c-y_b)}.
\end{equation} 
\end{subequations}  
Here, $\mathcal{N}_{t,x}=\cos{\vartheta_t}\cos{\alpha_t}$, $\mathcal{N}_{t,y}=\cos{\vartheta_t}\sin{\alpha_t}$, and $\mathcal{N}_{t,z}=\sin{\vartheta_t}$. 

Second, the energy contribution of each reflection region to the received energy $\mathcal{Q}_{r,\rm{ref}}$ is developed. When a pulse energy $\mathcal{Q}_t$ is transmitted by the transmitter, the unextinguished energy arriving at the differential reflection region d$\mathbb{U}$ can be derived as
\begin{equation} 
\mathcal{Q}_{{\rm{ref}},{\rm{d}}\mathbb{U}}=\frac{\mathcal{Q}_t \cos{\omega_i}\exp(-k_e \tau)}{2\pi(1-\cos{\beta_t})\tau^2}{\rm{d}}\mathbb{U},
\label{eq:dU}
\end{equation}
where $\omega_i$ is the incidence angle of the UV photon, which can be expressed as $\cos{\omega_i}=-\boldsymbol{n \tau}^{\rm{T}}/\tau$, and $\boldsymbol{n}$ denotes the normal vector of the reflection region. The Phong model is adopted~to describe the reflection pattern of the surface \cite{ref32}, which takes the diffuse and specular components into consideration. After the UV photons are reflected by ${\rm{d}}\mathbb{U}$, they arrive at the receiver aperture. The received energy contributed by ${\rm{d}}\mathbb{U}$ can be given by 
\begin{equation} 
{\rm{d}}\mathcal{Q}_{r,\rm{ref}}= r_r {\rm{I}}_r(\vartheta_1,\vartheta_2) \frac{A_r\cos{\vartheta_v}}{\varepsilon^{2}}\exp(-k_e\varepsilon) \mathcal{Q}_{{\rm{ref}},{\rm{d}}\mathbb{U}},
\label{eq:U}
\end{equation}
where $r_r$ is the reflection coefficient, $\vartheta_1$ and $\vartheta_2$ are the angles between $\boldsymbol{\varepsilon}$ and $\boldsymbol{n}$, and $\boldsymbol{\varepsilon}$ and $\boldsymbol{v_s}$, respectively, and $\boldsymbol{v_s}$ represents the direction vector of the specular reflection of $\boldsymbol{\tau}$. Moreover, ${\rm{I}}_r(\vartheta_1,\vartheta_2)$ can be expressed as \cite{ref20} 
\begin{equation}  
{\rm{I}}_r (\vartheta_1,\vartheta_2)=\eta\frac{\cos{\vartheta_1}}{\pi}+(1-\eta)\frac{m_s+1}{2\pi}\cos^{m_s}\vartheta_2,   
\end{equation}                
where $\eta$ is the percentage of the incident signal that is reflected diffusely and assumes values between $0$ and $1$, and $m_s$ is the directivity of the specular components. 

Substituting (\ref{eq:dU}) into (\ref{eq:U}) and integrating ${\rm{d}}\mathcal{Q}_{r,\rm{ref}}$ over these three reflection regions, the total received reflection energy can be expressed as
\begin{subequations}
\begin{equation}
\begin{aligned}
\mathcal{Q}_{r,\rm{ref}} & = \iint_{\mathbb{U}_{\rm{cdd'c'}}} r_r \mathcal{Q}_t {A_r} \mathcal{Z}\,\mathcal{Y}_{\rm{wei}}\,{\rm{d}}y{\rm{d}}z \\
                                            & + \iint_{\mathbb{U}_{\rm{daa'd'}}} r_r \mathcal{Q}_t {A_r} \mathcal{Z}\,\mathcal{Y}_{\rm{wei}}\,\mathcal{G}(\alpha)\,{\rm{d}}y{\rm{d}}z \\
                                            & + \iint_{\mathbb{U}_{\rm{bcc'b'}}} r_r \mathcal{Q}_t {A_r} \mathcal{Z}\,\mathcal{Y}_{\rm{wei}}\,\mathcal{B}(\alpha)\,{\rm{d}}y{\rm{d}}z,
\end{aligned}
\end{equation}
\begin{equation}  
\mathcal{Z}=\frac{{\rm{I}}_r (\vartheta_1,\vartheta_2) \cos{\vartheta_v} \cos{\omega_i} \exp[-k_e(\tau+\varepsilon)]}{2\pi(1-\cos{\beta_t})\tau^2 \varepsilon^2},
\end{equation} 
\end{subequations}
where
\begin{subequations}
\begin{equation}
\mathcal{Y}_{\rm{wei}}= 
\begin{cases}
1, & \beta_{t,\rm{esp}}\in[0,\beta_t], \\
    & \beta_{r,\rm{esp}}\in[0,\beta_r], \\
    & \rm{and\,\,Constraint}\,\mathbb{A},\\
0, & \rm{otherwise}.
\end{cases}
\end{equation}
\begin{equation}
\mathcal{G}(\alpha)= 
\begin{cases}
1, & \alpha<0,\\
0, & \alpha\ge0,
\end{cases}
\end{equation} 
\begin{equation}
\mathcal{B}(\alpha)= 
\begin{cases}
1, & \alpha>0,\\
0, & \alpha\le0,
\end{cases}
\end{equation}
\begin{equation}
\cos\vartheta_1=\frac{\boldsymbol{n}\boldsymbol{\varepsilon}^{\rm{T}}}{\varepsilon}, 
\end{equation}
\begin{equation}
\cos\vartheta_2=\left(\frac{\boldsymbol{\tau}}{\tau}-2\frac{\boldsymbol{\tau}}{\tau}\boldsymbol{n}^{\rm{T}}\boldsymbol{n}\right)\frac{\boldsymbol{\varepsilon}^{\rm{T}}}{\varepsilon},
\end{equation}
\end{subequations}
and $\boldsymbol{n}=[y_d-y_c,x_c-x_d,0]/w$ for $\mathbb{U}_{\rm{cdd'c'}}$ when $\alpha\ne0$, while when $\alpha=0$, $\boldsymbol{n}=[1,0,0]$. Regarding $\mathbb{U}_{\rm{daa'd'}}$, $\boldsymbol{n}$ can be expressed as $[y_a-y_d,x_d-x_a,0]/s$ when $\alpha<0$. As regards $\mathbb{U}_{\rm{bcc'b'}}$, $\boldsymbol{n}$ can be given by $[y_c-y_b,x_b-x_c,0]/s$ when $\alpha>0$. For $\mathbb{U}_{\rm{cdd'c'}}$ where $\alpha<0$ or $\mathbb{U}_{\rm{bcc'b'}}$ where $\alpha>0$, Constraint $\mathbb{A}$ denotes that $\Psi_{r,\rm{esp}}$ is less than $\Psi_{r,dd'}$ for $\mathbb{U}_{\rm{cdd'c'}}$ or less than $\Psi_{r,cc'}$ for $\mathbb{U}_{\rm{bcc'b'}}$, while for $\mathbb{U}_{\rm{cdd'c'}}$ where $\alpha>0$ or $\mathbb{U}_{\rm{daa'd'}}$ where $\alpha<0$, Constraint $\mathbb{A}$ represents that $\Psi_{t,\rm{esp}}$ is less than $\Psi_{t,cc'}$ for $\mathbb{U}_{\rm{cdd'c'}}$ or less than $\Psi_{t,dd'}$ for $\mathbb{U}_{\rm{daa'd'}}$. 

By summing the scattered energy derived in Section III and the reflected energy obtained in Section IV, the total received pulse energy can be expressed as
\begin{equation}
\mathcal{Q}_r=\mathcal{Q}_{r,\rm{sca}}+\mathcal{Q}_{r,\rm{ref}},
\end{equation} 
and the channel path loss for UV NLoS scenarios considering an obstacle can be expressed as
\begin{equation}
\mathcal{L}_{\rm{obs}}=10\log_{10}({\mathcal{Q}_t}/{\mathcal{Q}_r}).
\end{equation}  

\begin{figure}[t]  
\centering  
\includegraphics[scale=0.55]{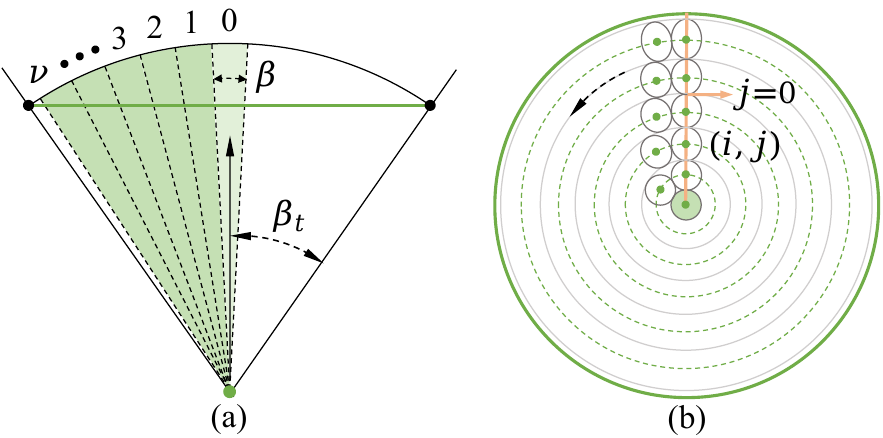}
\centering
\caption{Sampling of the transmitter beam: (a) the number of sampling layers for the entire beam and (b) the number of sampled sub-beams per layer.}
\label{Fig4}  
\end{figure}  

\section{Modeling of UV NLoS Propagation Channel Without Obstacles} 
To overcome the challenges faced by the existing SSSPLMs \cite{ref24}--\cite{ref28}, which apply to the cases where the overlap volume between the transceiver FoVs is small, or the transmitter beam (receiver FoV) angle is small, a high-accuracy SSSPLM is put forward in this section, which applies to cases with diversified transceiver FoV angles.

First, we sample the transmitter beam, as provided in Fig.~\ref{Fig4}, where $\beta$ is the sampling accuracy, and $\nu$ denotes the maximum number of sampling layers for the whole beam, which can be expressed as 
\begin{equation}
\nu=\left\lfloor\frac{\beta_t}{\beta}-\frac{1}{2}\right\rfloor.
\end{equation} 
On this basis, the number of sampled sub-beams for each layer, $\nu_i$, is investigated, where $i=0, 1, 2, ...,\nu$. For $i=0$, $\nu_0=1$, while for $i\ge1$, $\nu_i$ can be derived as 
\begin{equation}
\nu_{i}=\left\lfloor\frac{\pi}{\tan^{-1}\Upsilon(i)}\right\rfloor,
\end{equation}
and
\begin{subequations}
\begin{equation}
\Upsilon(i)=\frac{\sqrt{4\sin^2(i\beta)-4\cos^2\frac{\beta}{2}-\mathcal{A}}}{2\cos\frac{\beta}{2}},
\end{equation}
\begin{equation}
\mathcal{A}=\frac{\sin^2(2i\beta)}{\cos^2(i\beta)-\cos^2\frac{\beta}{2}}.
\end{equation}
\end{subequations}
Then, the energy contribution of each sub-beam to the received energy $\mathcal{Q}_r$ is determined.

Suppose that ${\rm{P}}_{i,j}$ is an arbitrary point on the sub-beam ${\rm{T}}_{i,j}$ axis and enclosed by the differential volume ${\rm{d}}v_{i,j}$, where ${\rm{d}}v_{i,j}$ is regarded as a thin layer of the spherical crown and can be expressed as 
\begin{equation}
{\rm{d}}v_{i,j}=2\pi(1-\cos\frac{\beta}{2})\tau^2_{i,j}{\rm{d}}\tau_{i,j},
\end{equation} 
and $j=0, 1, 2, ...,\nu_i-1$. $\tau_{i,j}$ and $\varepsilon_{i,j}$ are the distances from ${\rm{P}}_{i,j}$ to T and R, respectively. Referring to the single-scattering propagation theory \cite{ref26}, the received energy from ${\rm{d}}v_{i,j}$ can~be derived as
\begin{equation}
\begin{aligned}
{\rm{d}}\mathcal{Q}^{r}_{i,j}  = & \frac{\mathcal{Q}^{t}_{i,j} {\rm{P}}(\cos{\vartheta^s_{i,j}}) \cos{\vartheta^v_{i,j}}}{\varepsilon^2_{i,j}}\\
                                         & \times k_s A_r \exp[-k_e(\tau_{i,j}+\varepsilon_{i,j})] {\rm{d}} \tau_{i,j},
\end{aligned}
\label{eq:dQij}
\end{equation} 
where $\mathcal{Q}^{t}_{i,j}$ is the emitted pulse energy of the sub-beam, which can be derived as $\mathcal{Q}_t [1-\cos(\beta/2)]/(1-\cos\beta_t)$, $\vartheta^s_{i,j}$ and $\vartheta^v_{i,j}$ denote the angles between $\boldsymbol{{\rm{TP}}_{i,j}}$ and $\boldsymbol{{\rm{P}}_{i,j}}\boldsymbol{\rm{R}}$, and $\boldsymbol{{\rm{RP}}_{i,j}}$ and $\boldsymbol{\rm{RH}}$, respectively.

By integrating (\ref{eq:dQij}) over the effective scattering volume, the received energy $\mathcal{Q}^{r}_{i,j}$ contributed by the sub-beam ${\rm{T}}_{i,j}$ can be expressed as                 
\begin{equation}
\mathcal{Q}^{r}_{i,j}=\int_{\tau^{\min}_{i,j}}^{\tau^{\max}_{i,j}}f(\tau_{i,j})\,{\rm{d}} \tau_{i,j}, 
\label{Q-ij}
\end{equation}
where the expression of $f(\tau_{i,j})\,{\rm{d}} \tau_{i,j}$ is consistent with that of the right side in (\ref{eq:dQij}). Next, the expressions of $\tau^{\min}_{i,j}$ and $\tau^{\max}_{i,j}$ are developed.

If $\mathcal{S}_{1,i,j}=0$ and $\mathcal{S}_{2,i,j}\ne0$, $\tau^{\min}_{i,j}=\tau_{0,i,j}=-\mathcal{S}_{3,i,j}/\mathcal{S}_{2,i,j}$, and $\tau^{\max}_{i,j}=10\,\tau^{\min}_{i,j}$, where $\tau_{0,i,j}$ is assumed to be greater than $0$, and $\mathcal{S}_{1,i,j}$, $\mathcal{S}_{2,i,j}$, and $\mathcal{S}_{3,i,j}$ can be obtained via Appendix \ref{App:C}. If $\mathcal{S}_{1,i,j}\ne0$ and $\Delta_{i,j}=(\mathcal{S}_{2,i,j})^2-4\,\mathcal{S}_{1,i,j}\mathcal{S}_{3,i,j}\ge0$, $\tau_{1,i,j}$ and $\tau_{2,i,j}$ can be expressed as 
\begin{equation}
\tau_{1,i,j}=\min\left(\frac{-\mathcal{S}_{2,i,j}-\sqrt{\Delta_{i,j}}}{2\mathcal{S}_{1,i,j}},\frac{-\mathcal{S}_{2,i,j}+\sqrt{\Delta_{i,j}}}{2\mathcal{S}_{1,i,j}}\right),
\end{equation}
\begin{equation}
\tau_{2,i,j}=\max\left(\frac{-\mathcal{S}_{2,i,j}-\sqrt{\Delta_{i,j}}}{2\mathcal{S}_{1,i,j}},\frac{-\mathcal{S}_{2,i,j}+\sqrt{\Delta_{i,j}}}{2\mathcal{S}_{1,i,j}}\right),
\label{T-max}
\end{equation}
where $\tau^{\min}_{i,j}=\tau_{1,i,j}$ and $\tau^{\max}_{i,j}=\tau_{2,i,j}$ for the case that $\tau_{1,i,j}>0$, while when $\tau_{1,i,j}<0$ and $\tau_{2,i,j}>0$, $\tau^{\min}_{i,j}=\tau_{2,i,j}$ and $\tau^{\max}_{i,j}=10\,\tau^{\min}_{i,j}$.

According to the Gauss-Legendre quadrature rule \cite{ref28}, $\mathcal{Q}^{r}_{i,j}$ can be further derived as
\begin{align}
\mathcal{Q}^{r}_{i,j} & = \mathcal{E}_{i,j} \int_{-1}^{1} f (\mathcal{N}_{i,j}+\mathcal{E}_{i,j} t) {\rm{d}}t \notag \\
                                & \approx \mathcal{E}_{i,j} \sum_{q=1}^{u} w_{q} f (\mathcal{N}_{i,j}+\mathcal{E}_{i,j} t_q),     
\end{align}
where $\mathcal{E}_{i,j}=(\tau^{\max}_{i,j}-\tau^{\min}_{i,j})/2$, $\mathcal{N}_{i,j}=(\tau^{\max}_{i,j}+\tau^{\min}_{i,j})/2$, and $t_q$ is the $q$-th root of the Legendre polynomial $P_u(t)$, which can be expressed as         
\begin{equation}
P_u(t)=\frac{1}{2^u u!} \frac{{\rm{d}}^u}{{\rm{d}}t^u} (t^2-1)^u.  
\end{equation}
$P_u(t)$ has $u$ zeros in the interval $(-1,1)$, and the weight $w_q$ can be given by
\begin{equation}
w_q=\frac{2}{(1-t^2_q)[P'_u(t_q)]^2}.
\end{equation} 

By superimposing the energy contribution of each sub-beam to the received energy, $\mathcal{Q}_{r,\rm{sim}}$ can be derived as    
\begin{equation}
\mathcal{Q}_{r,\rm{sim}}=\sum_{i=0}^{\nu} \sum_{j=0}^{\nu_i-1} \sum_{q=1}^{u} \mathcal{E}_{i,j} w_{q} f (\mathcal{N}_{i,j}+\mathcal{E}_{i,j} t_q),
\end{equation}
and the corresponding path loss $\mathcal{L}_{\rm{sim}}$ can be expressed as 
\begin{equation}
\mathcal{L}_{\rm{sim}}=10\log_{10}\left(\sum_{i=0}^{\nu} \sum_{j=0}^{\nu_i-1}\frac{\mathcal{Q}^{t}_{i,j}}{\mathcal{Q}_{r,\rm{sim}}}\right).
\end{equation}  

\section{Numerical Results} 
In this section, we investigate the proposed NLoS scattering model, the proposed NLoS reflection model, and the proposed SSSPLM by comparing them with the related works.

\linespread{1.5}
\begin{table}[t]
\centering
\linespread{1.0}
\caption{Parameter Settings for Validation}
\label{t7}
\begin{tabular}{l l || l l}
\hline
\hline
{Parameters}&{Values}&{Parameters}&{Values}\\
\hline
\hline
$k^{\rm{Ray}}_s$&$0.24\,\rm{km^{-1}}$&$\beta_t$&$\pi/6$\\
$k^{\rm{Mie}}_s$&$0.25\,\rm{km^{-1}}$&$\beta_r$&$\pi/6$\\
$k_a$&$0.90\,\rm{km^{-1}}$&$\alpha_t$&$19\pi/36$\\
$\gamma$&$0.017$&$\alpha_r$&$-19\pi/36$\\
g&0.72&$A_r$&$1.92\,\rm{cm}^2$\\
$f$&0.5&$m_s$&5\\
$r_r$&0.1&$\eta$&0.5\\
\hline
\hline
\end{tabular}
\end{table}
\linespread{1.0} 

\begin{figure}[t]  
\centering  
\includegraphics[scale=0.42]{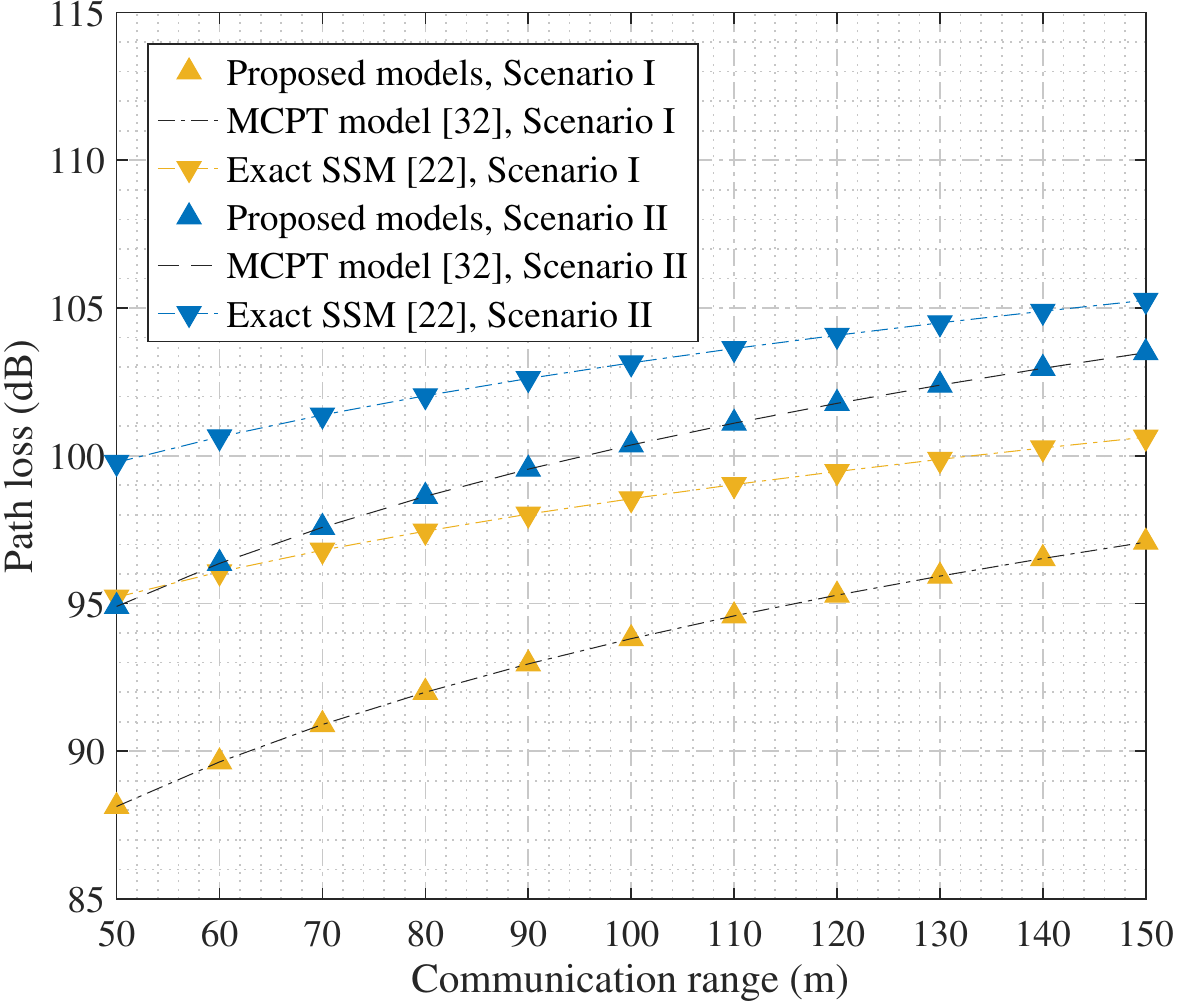}
\centering
\caption{Path loss results for the proposed model, the MCPT model \cite{ref32}, and the integration model \cite{ref22} under different scenarios.}
\label{Fig5}  
\end{figure}

First, the proposed NLoS scattering model and the proposed NLoS reflection model are validated by comparing them to the latest Monte-Carlo photon-tracing (MCPT) model \cite{ref32}, which tracks the propagation process of numerous photons to obtain the stable statistical results at the receiver. As shown in Fig.~\ref{Fig5}, the obstacle parameters $s$, $w$, $\kappa$, $x_o$, $y_o$, and $\alpha$ are set to $r/10$, $2r$, $2r$, $x_{o,\max}-s$, $r/2$, and $0^{\circ}$, respectively, and the reflection parameters $r_r$, $m_s$, and $\eta$ of the reflection surface are selected as $0.1$, $5$, and $0.5$, respectively, which are consistent with the parameter settings in~\cite{ref32}. As regards the remaining parameter settings, they are summarized in Table~\ref{t7}, where the selected transceiver elevation, azimuth, and FoV angles can ensure that the overlap volume between the transmitter beam and receiver FoV intersects with the obstacle, which is helpful to the study of the contribution of obstacle reflection to the received pulse energy. Besides, the settings of atmospheric model parameters are in accordance with existing works \cite{ref24,ref32}. From Fig.~\ref{Fig5} it can be found that the path loss curves obtained by the proposed models coincide well with those obtained by the MCPT model under different scenarios, which implies that the accuracies of the proposed NLoS scattering model and reflection model are comparable to that of the MCPT model. Here, Scenarios I and II represent that $\vartheta_t$ and $\vartheta_r$ are set to $25^{\circ}$ and $35^{\circ}$, respectively. For the simulation time, the MCPT model is much longer than the proposed scattering and reflection models. Specifically, in the computing environment of MATLAB R2022b installed on a computer with 2 GHz Quad-Core Intel Core i5 and 16 GB LPDDR4X, the simulation time of the MCPT model is at least one order of magnitude longer than that of the proposed model. During the entire simulation, the number of collision-induced events in \cite{ref32} is set to one because the single-scattering events are dominant in short-range UV NLoS scenarios. Furthermore, the number of simulated photons and their survival probability threshold are set to tens of millions and $10^{-10}$, respectively, which can ensure that the simulation error of the MCPT model is less than 0.1 dB.

Then, the path loss for UV NLoS communication scenarios with and without obstacles incorporated is investigated under equivalent parameter settings. From Fig.~\ref{Fig5} it can be found that obstacle reflections can improve the channel path loss for UV NLoS communication systems. For example, when the range is set to 100 m for Scenario I, the path loss determined by the proposed models and the path loss obtained by the integration model \cite{ref22} (no obstacle in communication scenarios) are 93.81 dB and 98.55 dB, respectively. These results demonstrate that avoiding obstacles is not always a good option in this case.  

\begin{figure}[t]  
\centering  
\includegraphics[scale=0.42]{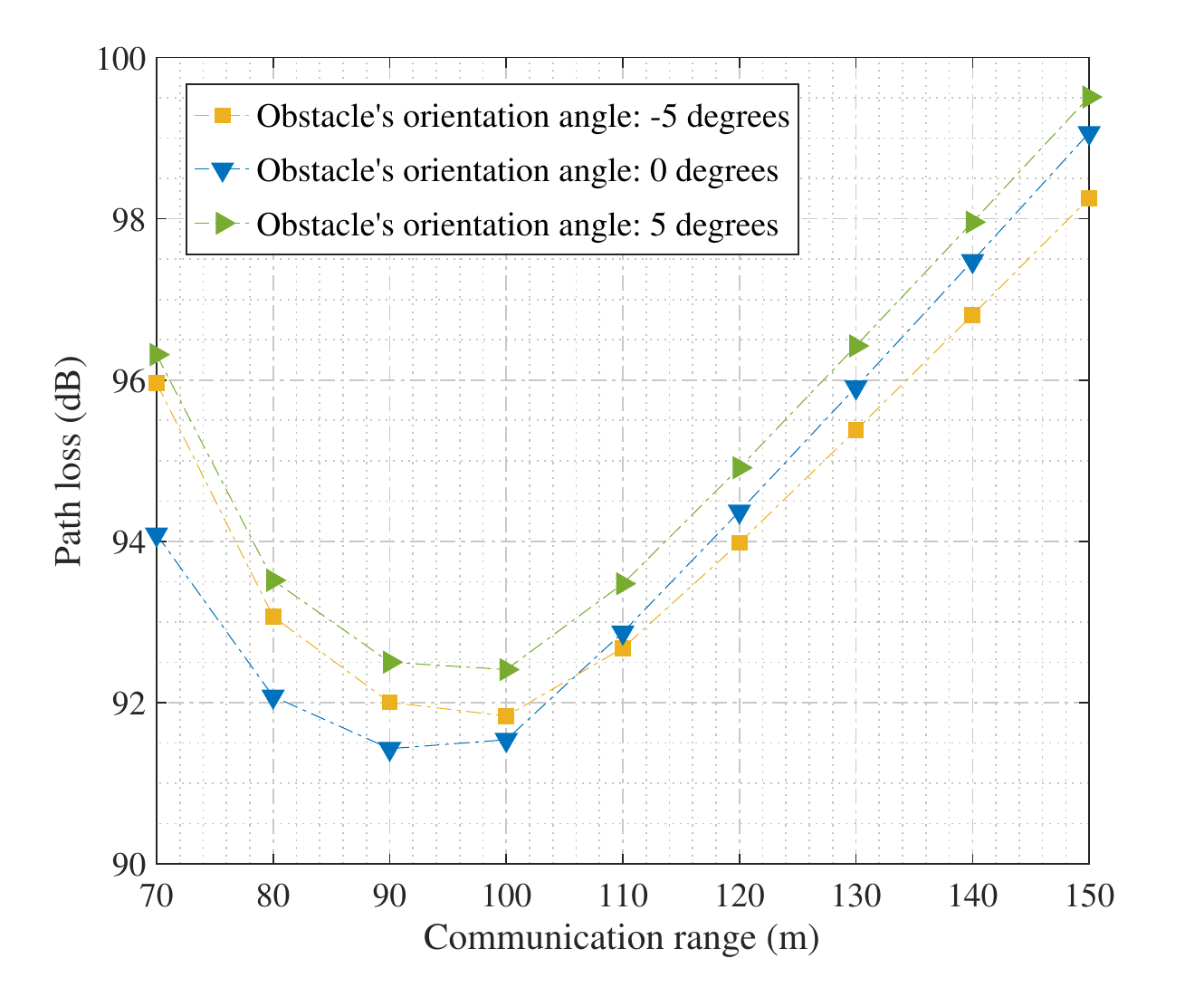}
\centering
\caption{Relationship between path loss and communication range for different obstacle orientation angles.}
\label{Fig6}  
\end{figure}

Further, the effects of the obstacle's orientation angle on UV channel path loss are investigated. To facilitate reproducing the numerical results, relevant parameter settings are summarized in Table~\ref{t8}. Fig.~\ref{Fig6} manifests that when the orientation angle of the obstacle remains unchanged, the channel path loss first decreases and then increases as $r$ increases. This is primarily because when $r$ is small, the adverse effects of the obstacle on the effective scattering volume and effective reflection regions are relatively obvious, which results in a large path loss. While as $r$ increases, these adverse effects gradually weaken, which leads to a gradual decrease in path loss. Then, as $r$ increases to a certain degree, the influence of $r$ on UV signals begins to dominate, which causes the path loss to start increasing again. In addition to this, an obvious phenomenon that can be found in Fig.~\ref{Fig6} is that when $r$ is smaller than or equal to 100 m, the path loss corresponding to 0 degrees is the smallest, while the path loss corresponding to 5 degrees is the largest at the same $r$ value. However, when $r$ is greater than 100 m, the path loss corresponding to $-5$ degrees becomes the minimum value at the same $r$ value. The main reason for this phenomenon is that the variation of $r$ and the obstacle's orientation angle alters the effective regions of the possible reflection surfaces. Moreover, the influence on the path loss when the area of the beam at the obstacle (ABO) is much greater than the area of the obstacle available for reflection (AOAR) is analyzed. Specifically, when the overall size of the obstacle is small, the adverse impact of ABO being much greater than AOAR on the path loss can be neglected, while when the overall size of the obstacle is large, this adverse impact can not be neglected because the effective scattering volume is reduced obviously.

\linespread{1.5}
\begin{table}[t]
\centering
\linespread{1.0}
\caption{Parameter Settings for Transceiver and Obstacle}
\label{t8}
\begin{tabular}{l l || l l}
\hline
\hline
{Parameters}&{Values}&{Parameters}&{Values}\\
\hline
\hline
$\beta_t$&$\pi/12$&$w$&40 m\\
$\beta_r$&$\pi/12$&$\kappa$&80 m\\
$\alpha_t$&$2\pi/3$&$s$&30 m\\
$\alpha_r$&$-2\pi/3$&$x_o$&$x_{o,\max}-s$\\
$\vartheta_t$&$\pi/9$&$y_o$&$r/2$\\
$\vartheta_r$&$\pi/9$&$\alpha$&$-5^{\circ},\,0^{\circ},\,5^{\circ}$\\
\hline
\hline
\end{tabular}
\end{table}
\linespread{1.0} 

\begin{figure}[t]  
\centering  
\includegraphics[scale=0.42]{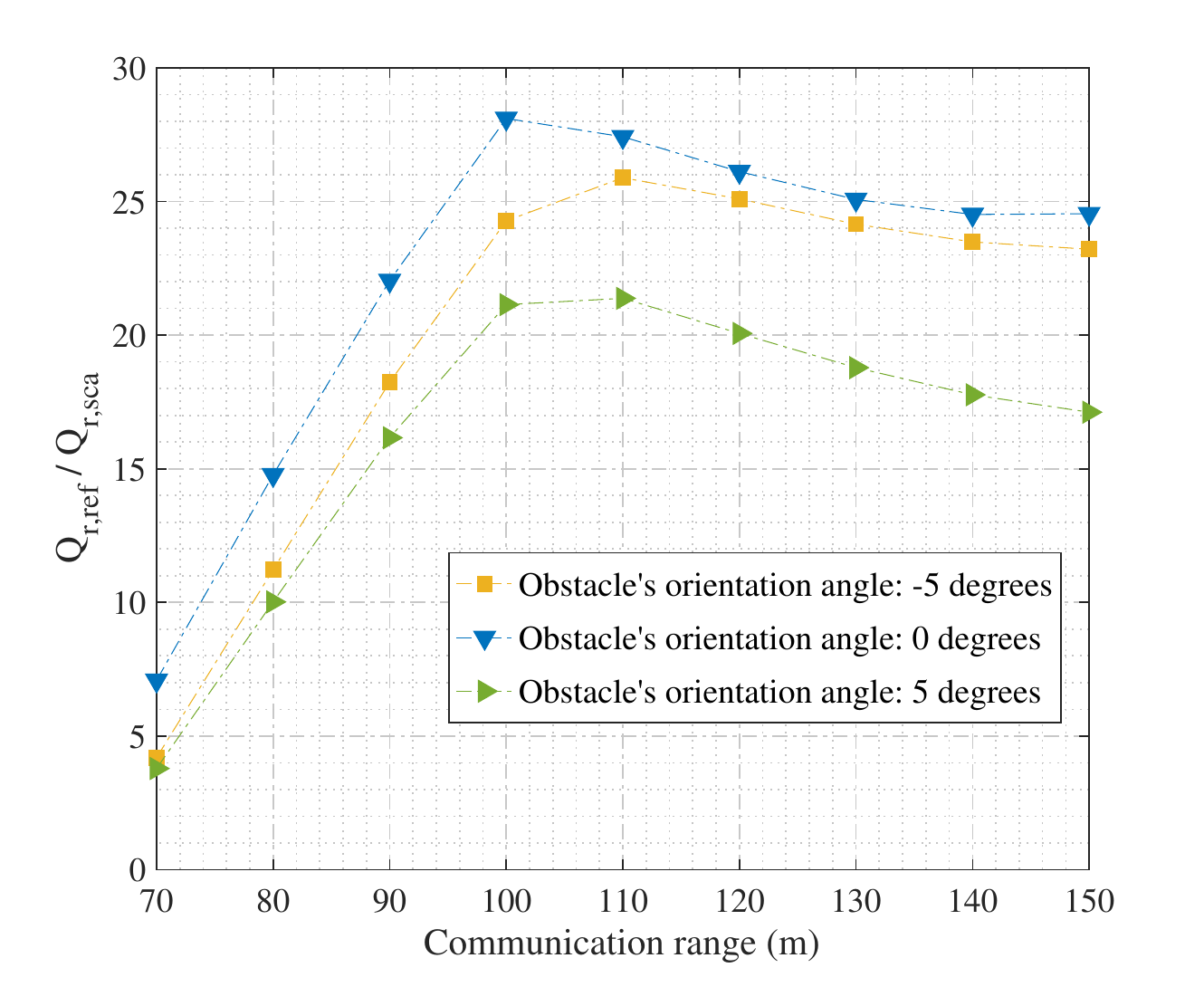}
\centering
\caption{Relationship between $\mathcal{Q}_{r,\rm{ref}}/\mathcal{Q}_{r,\rm{sca}}$ and communication range versus different obstacle orientation angles.}
\label{Fig7}  
\end{figure}

\begin{figure}[t]  
\centering  
\includegraphics[scale=0.42]{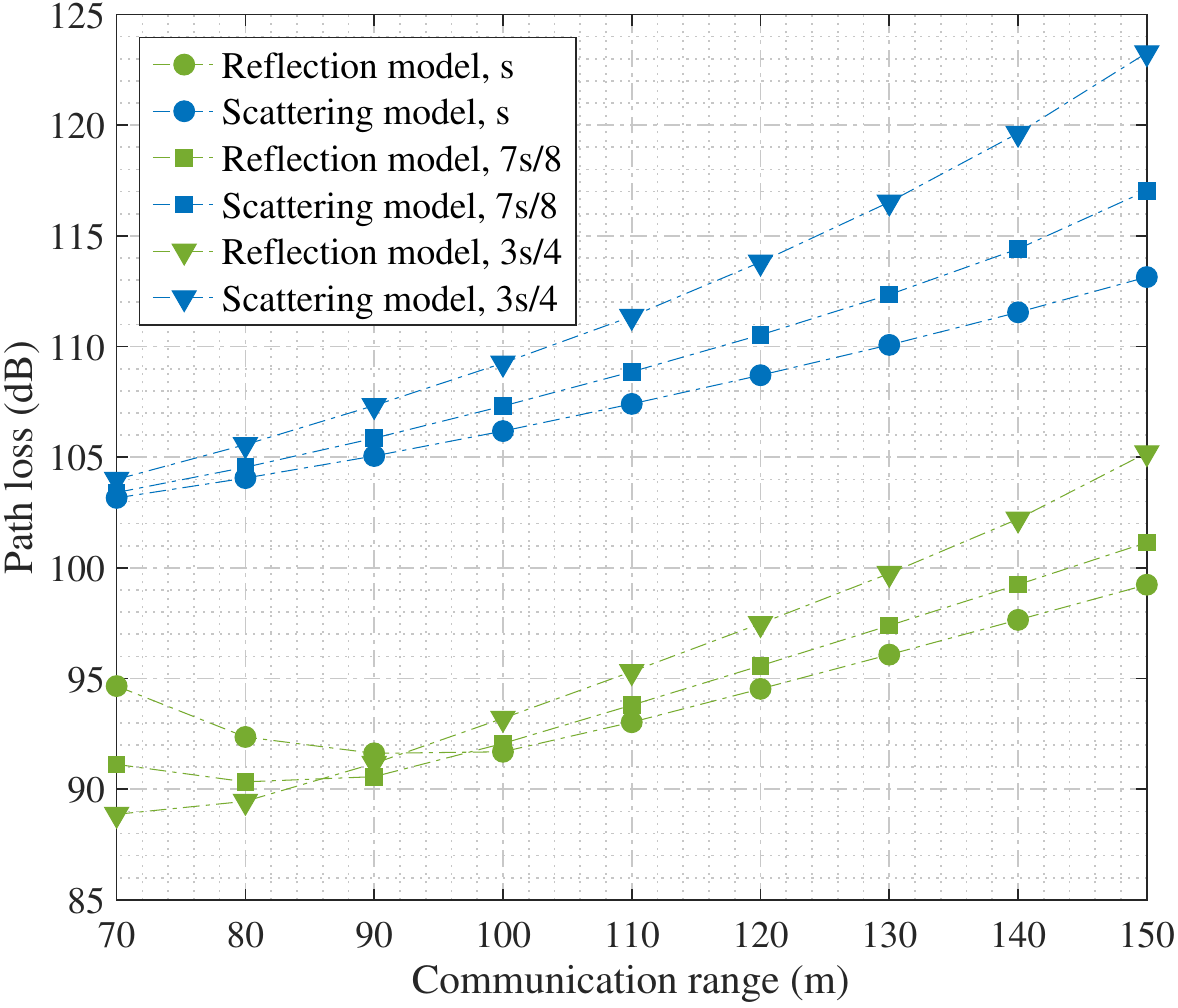}
\centering
\caption{Path loss determined by the scattering model and the reflection model versus different $x_o$ values.}
\label{Fig8}  
\end{figure}

Moreover, the contributions of $\mathcal{Q}_{r,\rm{sca}}$ and $\mathcal{Q}_{r,\rm{ref}}$ to the total received energy $\mathcal{Q}_{r}$ are analyzed. From Fig.~\ref{Fig7}, it can be found that under given parameter settings, the contribution of $\mathcal{Q}_{r,\rm{ref}}$ to $\mathcal{Q}_{r}$ is clearly more than that of $\mathcal{Q}_{r,\rm{sca}}$ to $\mathcal{Q}_{r}$. For example, when $r$ is set to 100 m, the values of $\mathcal{Q}_{r,\rm{ref}}/\mathcal{Q}_{r,\rm{sca}}$ for cases $-5$ degrees, 0 degrees, and 5 degrees are 24.28, 28.11, and 21.14, respectively. Then, the impacts of the distance between the reflection surface and the Y axis (i.e., $x_{o,\max}-x_o$) on the reflected energy and the scattered energy are investigated, as presented in Fig.~\ref{Fig8}, where $\alpha$ is set to 0 degrees. The numerical results manifest that as the value of $x_{o,\max}-x_o$ increases, the path loss of the reflected energy is overall lower than that of the scattered energy. These findings suggest that in the actual application scenarios, we should utilize obstacle reflections as much as possible to reduce the path loss of the NLoS channel, thereby improving system communication performance.


\begin{figure}[t]  
\centering  
\includegraphics[scale=0.42]{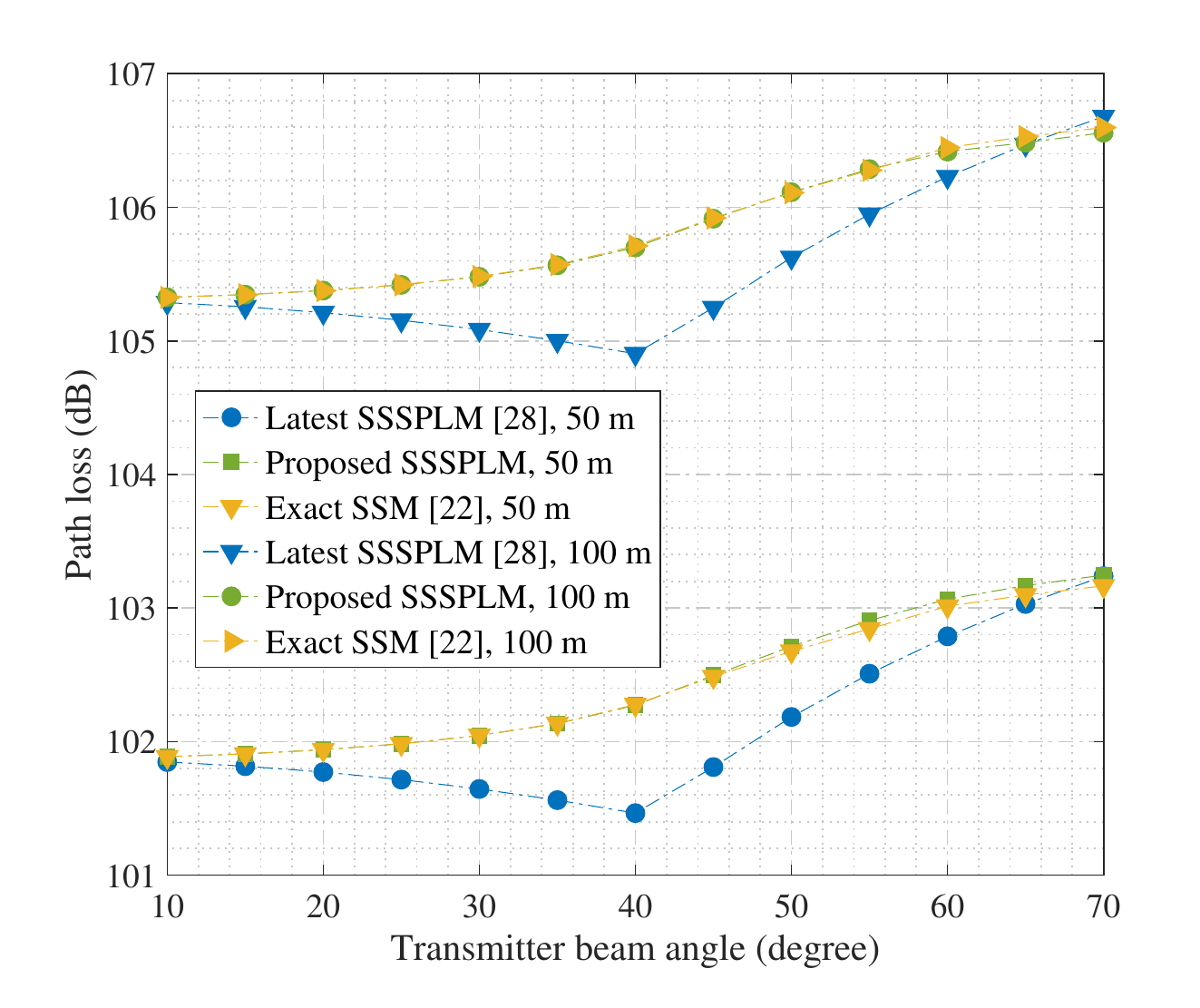}
\centering
\caption{Path loss versus the transmitter beam angle for the proposed SSSPLM, the exact SSM \cite{ref22}, and the latest SSSPLM \cite{ref28}.}
\label{Fig9} 
\end{figure} 
\begin{figure}[t]  
\centering  
\includegraphics[scale=0.42]{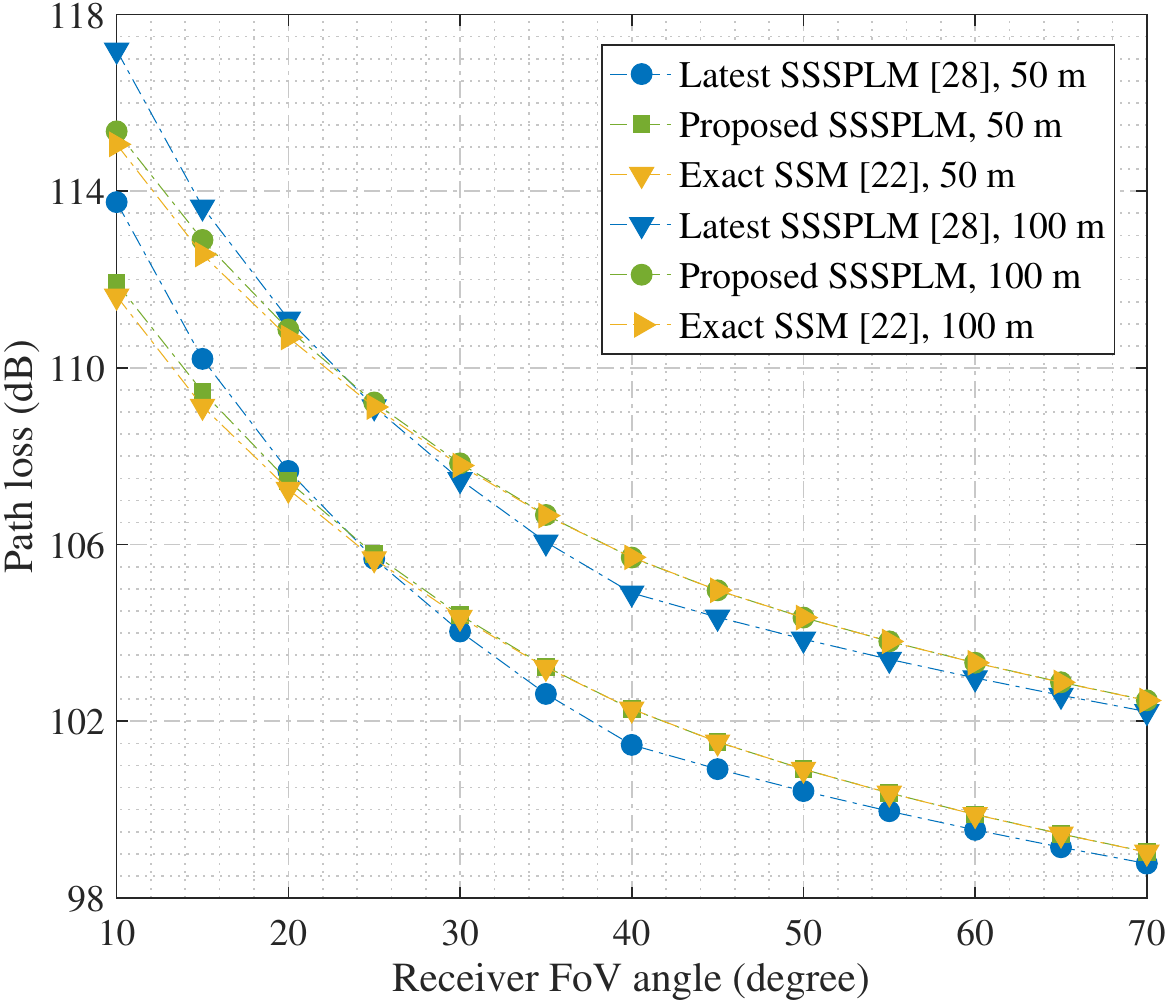}
\centering
\caption{Path loss versus the receiver FoV angle for the proposed SSSPLM, the exact SSM \cite{ref22}, and the latest SSSPLM \cite{ref28}.}
\label{Fig10}  
\end{figure}

In addition, we verify the accuracy of the proposed SSSPLM by comparing it with the exact SSM~\cite{ref22} and the state-of-the-art SSSPLM~\cite{ref28}. First, the accuracy of the proposed SSSPLM is investigated with the variation of transceiver FoV angles, as shown in Figs.~\ref{Fig9} and~\ref{Fig10}, where system model parameters are selected as follows: $\vartheta_t$ and $\vartheta_r$ are both set to $\pi/4$; $\alpha_t$ and $\alpha_r$ are set to $19\pi/36$ and $-19\pi/36$, respectively; $\beta_r$ is set to $\pi/9$ for Fig.~\ref{Fig9}; $\beta_t$ is set to $\pi/9$ for Fig.~\ref{Fig10}; the sampling accuracy is set to $\beta_t/50$; and $u$ is set to 30. Calculation results manifest that the proposed SSSPLM agrees better with the exact SSM than the latest SSSPLM. To quantitatively assess the accuracy of the proposed model, the root-mean-square error (RMSE) is introduced \cite{ref28} and can be expressed as 
\begin{equation}
{\rm{RMSE}}=\sqrt{\frac{1}{\mathcal{M}}\sum_{i=1}^{\mathcal{M}}\left[\mathcal{L}_{\rm{exa}}(x_i)-\mathcal{L}_{\rm{sim}}(x_i)\right]^2},
\end{equation}
where $x_i$ denotes the $i$-th ($i=1, 2, ..., \mathcal{M}$) value of the given variable, which could be the transmitter beam angle (i.e., $2\beta_t$), the receiver FoV angle (i.e., $2\beta_r$), and the transceiver elevation angles. $\mathcal{L}_{\rm{exa}}(x_i)$ denotes the channel path loss (unit: dB) of $x_i$ obtained by the exact model, and $\mathcal{L}_{\rm{sim}}(x_i)$ denotes the path loss (unit: dB) of $x_i$ obtained by the proposed simplified model or the latest simplified model. For example, when $r$ is set to 50 m in Fig.~\ref{Fig9}, the RMSEs of the proposed simplified model and the latest simplified model are 0.04~dB and 0.41~dB, respectively, while when $r$ is set to 100 m, the counterparts are 0.02 dB and 0.40 dB, respectively. Regarding Fig.~\ref{Fig10}, when $r$ is set to 50 m or 100m, the RMSEs of the proposed simplified model and the latest simplified model are 0.14 dB and 0.79~dB, respectively.

\begin{figure}[t]  
\centering  
\includegraphics[scale=0.42]{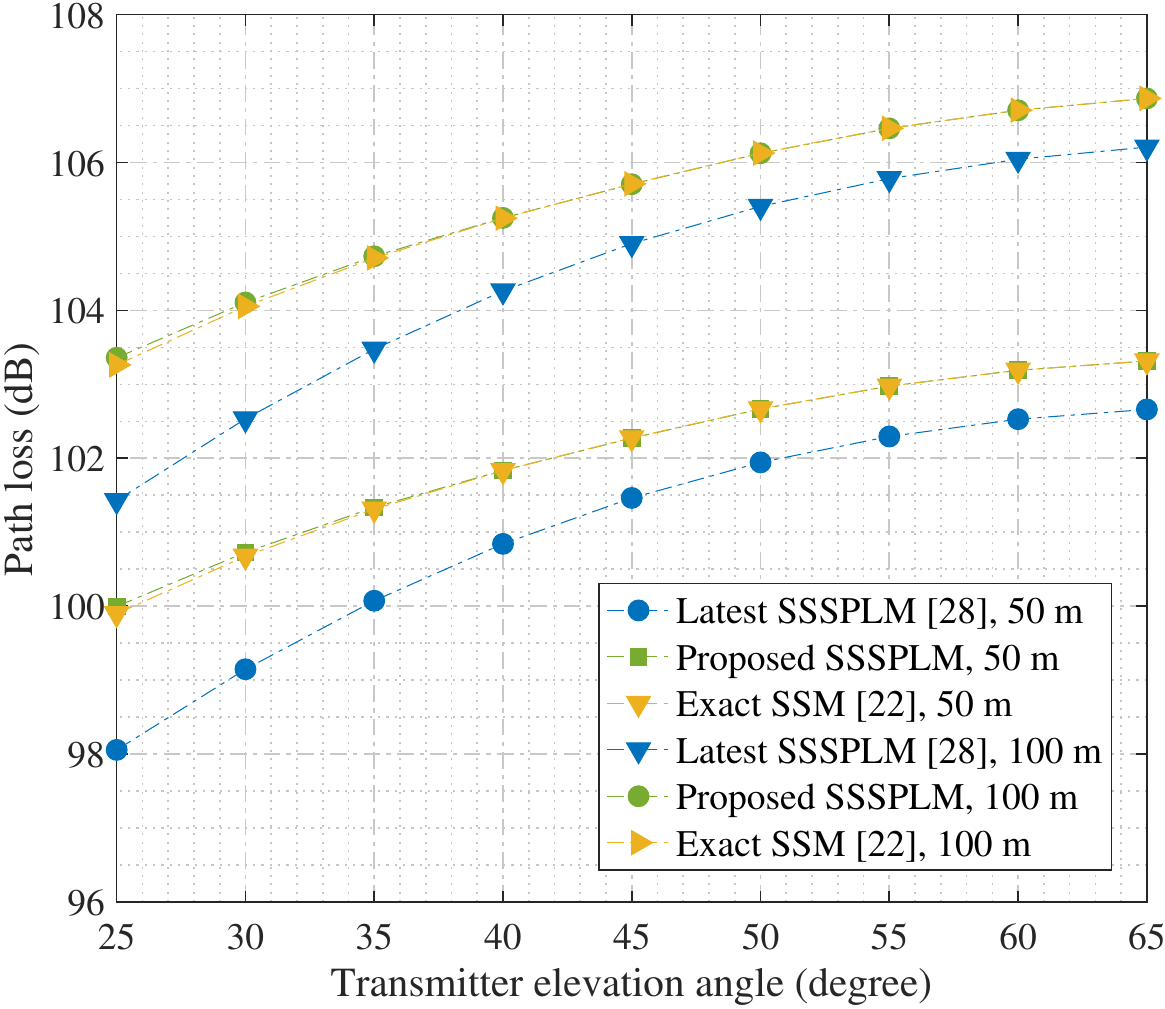}
\centering
\caption{Path loss versus the transmitter elevation angle for the proposed SSSPLM, the exact SSM \cite{ref22}, and the latest SSSPLM \cite{ref28}.}
\label{Fig11}  
\end{figure}
\begin{figure}[t]  
\centering  
\includegraphics[scale=0.42]{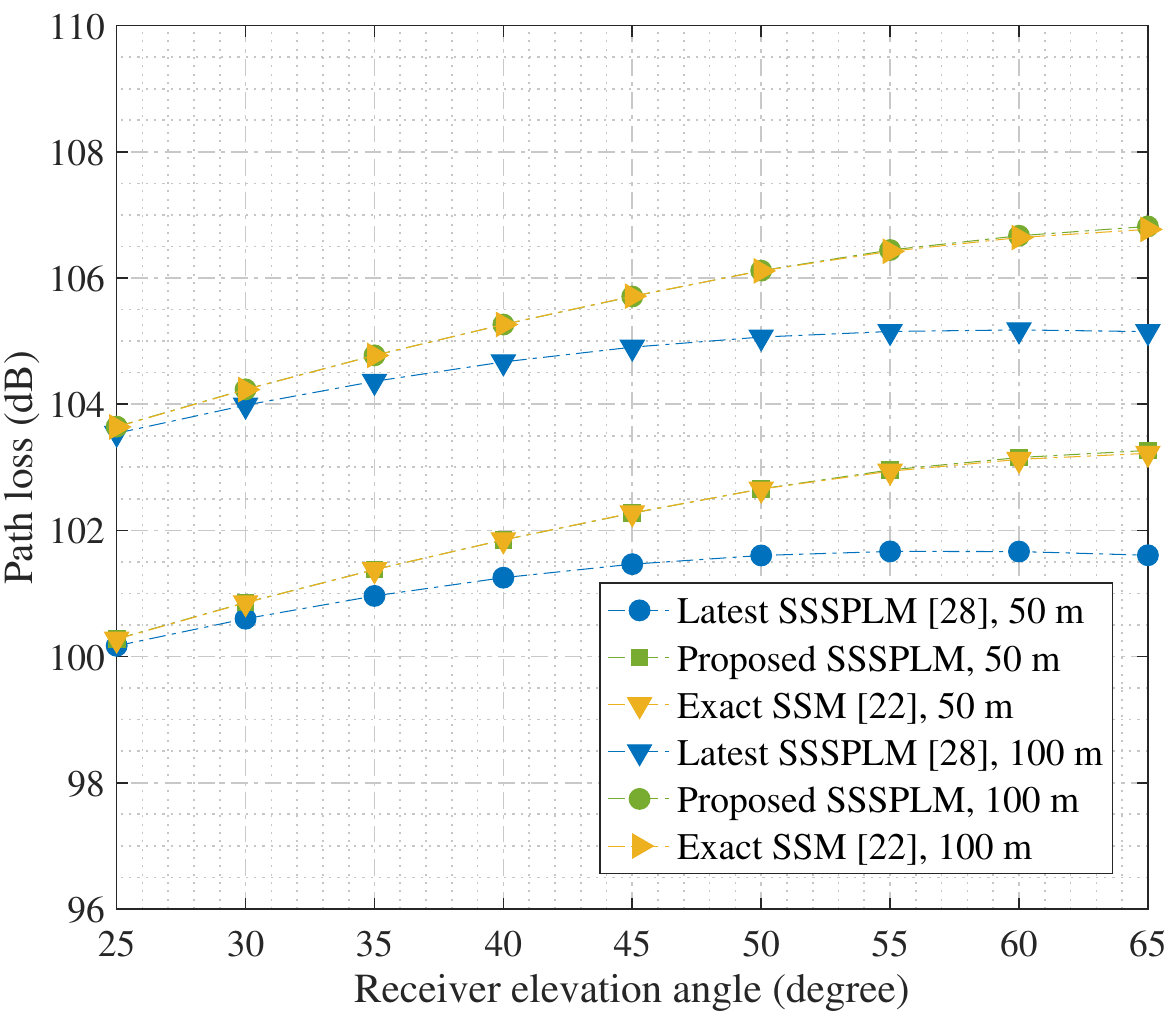}
\centering
\caption{Path loss versus the receiver elevation angle for the proposed SSSPLM, the exact SSM \cite{ref22}, and the latest SSSPLM \cite{ref28}.}
\label{Fig12}  
\end{figure}   

Finally, the accuracy of the proposed SSSPLM is examined with the variation of transceiver elevation angles, as shown in Figs.~\ref{Fig11} and~\ref{Fig12}, where $\beta_t$ and $\beta_r$ are both set to $\pi/9$, and $\vartheta_r$ for Fig.~\ref{Fig11} and $\vartheta_t$ for Fig.~\ref{Fig12} are both set to $\pi/4$. Calculation results show that the proposed SSSPLM agrees better with the exact SSM than the state-of-the-art SSSPLM. Specifically, when $r$ is set to 50 m or 100 m in Fig.~\ref{Fig11}, the RMSEs of the proposed simplified model and the latest simplified model are 0.04~dB and 1.09 dB, respectively, while when $r$ is set to 50 m or 100~m in Fig.~\ref{Fig12}, the counterparts are 0.02 dB and 0.99~dB, respectively. Based on these results, it can be found that the proposed SSSPLM performs very well in evaluating the channel attenuation of UV NLoS communication systems incorporating various transceiver FoV/elevation angles, and its accuracy can be further improved by increasing the sampling accuracy $\beta$ and the roots $u$ of the Legendre polynomial. Note that although we investigate the performance of the proposed models under the given system parameters such as the obstacle reflection coefficient as well as the atmospheric scattering and absorption coefficients, they are also applicable to other values of system parameters. Because the correctness of their physical modeling mechanisms is validated by the related works under identical parameter settings.

\section{Conclusion} 
In this paper, we proposed a UV NLoS scattering model and a reflection model incorporating an obstacle. To approach real communication environments, the obstacle's orientation angle, coordinates, and geometric dimensions were considered during this modeling process. Besides, due to the high computational complexity of the existing exact SSMs without obstacles, and the limited applicability of the existing SSSPLMs, an SSSPLM was developed with a closed-form expression. Following that, the proposed models were validated by comparing them with the MCPT model, the exact SSM, and the latest SSSPLM. The numerical results show that the path loss curves determined by the proposed models agree well with those obtained from the related works under identical parameter settings. Moreover, the proposed SSSPLM performs better than the latest SSSPLM in assessing the channel attenuation of UV NLoS communication systems considering various transceiver elevation/FoV angles, and its accuracy can be improved by increasing the sampling accuracy $\beta$ and the root number of the Legendre polynomial. This work disclosed that when the intersection area among the transmitter beam, the receiver FoV, and the area of the obstacle available for reflection is relatively large, the channel path loss of UV NLoS systems considering obstacles is obviously lower than that of UV NLoS systems without any obstacle, and the energy contribution of obstacle reflections to the whole signal is distinctly higher than that of atmospheric scattering, which inspires us to make more use of obstacle reflections to improve the communication performance of UV NLoS systems.

\begin{appendices} 

\section{Derivation of $|J_3|$}
\label{App:A}
Suppose that $\rm{P}$ is located on the plane $\mathcal{F}_{\vartheta}$ and enclosed by the transmitter beam, its coordinates can be expressed as  
\begin{equation}
{\rm{P}:}
\begin{cases}
x=\tau\cos{\varpi}\cos{\delta_t}\,{\cos(\alpha_t+\phi)}{\sec{\phi}},\\
y=\tau\cos{\varpi}\cos{\delta_t}\,{\sin(\alpha_t+\phi)}{\sec{\phi}},\\
z=\tau\cos{\varpi}\sin{\delta_t},
\end{cases}
\label{eq:Pzb}  
\end{equation}
where $\delta_t=\vartheta_t+\vartheta$ and $\phi=\tan^{-1}(\tan{\varpi}\sec{\delta_t})$.    

By transforming the differential volume from the cartesian coordinate system ${\rm{d}}x{\rm{d}}y{\rm{d}}z$ to the spherical coordinate system $|J_3|{\rm{d}}\tau{\rm{d}}\varpi{\rm{d}}\vartheta$, $|J_3|$ can be determined via Jacobian determinant and can be derived as
\begin{align}
|J_3| = & \,\frac{\partial x}{\partial \tau}\left(\frac{\partial y}{\partial \varpi}\frac{\partial z}{\partial \vartheta}-\frac{\partial y}{\partial \vartheta}\frac{\partial z}{\partial \varpi}\right)+\frac{\partial x}{\partial \varpi}\left(\frac{\partial y}{\partial \vartheta}\frac{\partial z}{\partial \tau}-\frac{\partial y}{\partial \tau}\frac{\partial z}{\partial \vartheta}\right)\nonumber\\
                    +& \,\frac{\partial x}{\partial \vartheta}\left(\frac{\partial y}{\partial \tau}\frac{\partial z}{\partial \varpi}-\frac{\partial y}{\partial \varpi}\frac{\partial z}{\partial \tau}\right)=\tau^2\cos{\varpi}.
\end{align} 

\section{Derivation of $\Psi_{r,\min}$ and $\Psi_{r,\max}$}
\label{App:B}
Since the values of $\Psi_{r,\rm{min}}$ and $\Psi_{r,\rm{max}}$ are closely related to the value of $\delta_r=\vartheta_r+\sigma$, we will deduce them from the following four intervals:  

$\boldsymbol{\delta_r\in[\delta_{r,\rm{low}},\Theta_{r,b}]}$: In this case, $\Psi_{r,\rm{min}}$ and $\Psi_{r,\rm{max}}$ can be derived as 
\begin{subequations}
\begin{equation}
\Psi_{r,\rm{min}}=\min(\Psi_{r,bb'},\Psi_{r,cc'},\Psi_{r,dd'}),
\end{equation} 
\begin{equation}
\Psi_{r,\rm{max}}=\max(\Psi_{r,aa'},\Psi_{r,bb'},\Psi_{r,dd'}),
\end{equation} 
\end{subequations}
where the angle $\Psi_{r,nn'}$ between $\boldsymbol{{\rm{RP}}_{r, nn'}}$ and $\boldsymbol{\rm{RS}}$ can be expressed as
\begin{equation}
\Psi_{r,nn'}=\cos^{-1}\left(\frac{r\Xi_{r,c}-y_{nn'}\Xi_{r,c}+z_{r,nn'}\Xi_{r,b}}{\sqrt{\Xi^2_{r,b}+\Xi^2_{r,c}}||\boldsymbol{{\rm{RP}}_{r,nn'}}||}\right),
\label{eq:Psir} 
\end{equation} 
and the subscript $n$ can be $a$, $b$, $c$ or $d$. The Z coordinate of ${\rm{P}}_{r,nn'}$ can be given by 
\begin{equation}
z_{r,nn'}=(r\Xi_{r,b}-y_n \Xi_{r,b}-x_n \Xi_{r,a})/\Xi_{r,c},
\end{equation} 
where $\Xi_{r,a}$, $\Xi_{r,b}$, and $\Xi_{r,c}$ can be expressed as 
\begin{subequations} 
\begin{equation}
\Xi_{r,a}=-{\cos{\alpha_r}\sec^2{\sigma}\sin(2\delta_r)\tan{\varphi_r}},
\end{equation} 
\begin{equation}
\Xi_{r,b}=-{\sin{\alpha_r}\sec^2{\sigma}\sin(2\delta_r)\tan{\varphi_r}},
\end{equation} 
\begin{equation}
\Xi_{r,c}={2\sec^2{\sigma}\cos^2\delta_r\tan{\varphi_r}},
\end{equation} 
\end{subequations}
and $\varphi_r=\tan^{-1}[\sec{\delta_r}(\tan^2\beta_r-\tan^2\sigma)^{1/2}\cos{\sigma}]$.

$\boldsymbol{\delta_r\in(\Theta_{r,b},\Theta_{r,c}]}$: In this case, $\Psi_{r,\rm{min}}$ and $\Psi_{r,\rm{max}}$ can be derived as 
\begin{subequations}
\begin{equation}
\Psi_{r,\rm{min}}=\min(\Psi_{r,bc},\Psi_{r,cc'},\Psi_{r,dd'}),
\end{equation} 
\begin{equation}
\Psi_{r,\rm{max}}=\max(\Psi_{r,aa'},\Psi_{r,ab},\Psi_{r,dd'}),
\label{eq:Prmax} 
\end{equation} 
\end{subequations} 
where $\Psi_{r,ab}$ and $\Psi_{r,bc}$ can be obtained through (\ref{eq:Psir}), e.g., by replacing the subscript $n$ with $a$ ($b$) and $n'$ with $b$ ($c$), and the $x_{r,mn}$ and $y_{r,mn}$ coordinates of ${\rm{P}}_{r, m n}$ can be given by
\begin{subequations}
\label{eq:Prmn}
\begin{align}
&x_{r,mn}=(y_{r,mn}-y_n)/\kappa_{r,mn}+x_n,\\ 
&y_{r,mn}=\frac{y_{n}\Xi_{r,a}+\kappa_{r,mn}(r\Xi_{r,b}-\kappa\Xi_{r,c}-x_n\Xi_{r,a})}{\Xi_{r,a}+\kappa_{r,mn}\Xi_{r,b}},
\end{align} 
\end{subequations} 
where $\kappa_{r,mn}=(y_m-y_n)/(x_m-x_n)$ under the conditions that $x_m \neq x_n$ and $y_m \neq y_n$, while if $x_m=x_n$ and $y_m \neq y_n$, $y_{r,mn}=(r\Xi_{r,b}-\kappa\Xi_{r,c}-x_m\Xi_{r,a})/\Xi_{r,b}$, and if $x_m \neq x_n$ and $y_m=y_n$, $x_{r,mn}=(r\Xi_{r,b}-\kappa\Xi_{r,c}-y_m \Xi_{r,b})/\Xi_{r,a}$. Moreover, if $x_m=x_n$ and $y_m=y_n$, ${\rm{P}}_{r, m n}$ is point C. 

$\boldsymbol{\delta_r\in(\Theta_{r,c},\Theta_{r,a}]}$: In this case, $\Psi_{r,\rm{min}}$ can be derived as 
\begin{equation}
\Psi_{r,\rm{min}}=\min(\Psi_{r,cd},\Psi_{r,dd'}),
\label{eq:min}
\end{equation} 
where ${\rm{P}}_{r, c d}$ can be obtained through (\ref{eq:Prmn}a)$\sim$(\ref{eq:Prmn}b), and $\Psi_{r,\rm{max}}$ is in accordance with (\ref{eq:Prmax}).

$\boldsymbol{\delta_r\in(\Theta_{r,a},\Theta_{r,d})}$: In this case, $\Psi_{r,\rm{max}}$ can be given by  
\begin{equation}
\Psi_{r,\rm{max}}=\max(\Psi_{r,ad},\Psi_{r,dd'}),
\end{equation}
where ${\rm{P}}_{r, a d}$ can be obtained through (\ref{eq:Prmn}a)$\sim$(\ref{eq:Prmn}b), and $\Psi_{r,\rm{min}}$ is in accordance with (\ref{eq:min}). 

\section{Derivation of $\mathcal{S}_{1,i,j}$, $\mathcal{S}_{2,i,j}$, and $\mathcal{S}_{3,i,j}$}
\label{App:C} 
Combining the equation of the sub-beam axis ${\rm{T}}_{i,j}$ with that of the receiver conical surface, $\mathcal{S}_{1,i,j}$, $\mathcal{S}_{2,i,j}$, and $\mathcal{S}_{3,i,j}$ can be derived as   
\begin{subequations}
\begin{equation}
\mathcal{S}_{1,i,j}=(\mathcal{N}_{r,x} \mathcal{A}_{i,j}+\mathcal{N}_{r,y} \mathcal{B}_{i,j}+\mathcal{N}_{r,z} \mathcal{C}_{i,j})^2-\cos^2{\beta_r},
\end{equation}
\begin{equation}
\begin{aligned}
\mathcal{S}_{2,i,j} = & 2\,r(\cos^2{\beta_r}\,\mathcal{B}_{i,j}-\mathcal{N}^2_{r,y}\,\mathcal{B}_{i,j}-\mathcal{N}_{r,x} \mathcal{N}_{r,y} \mathcal{A}_{i,j}-\\
                                 & \mathcal{N}_{r,y} \mathcal{N}_{r,z} \mathcal{C}_{i,j}), 
\end{aligned}
\end{equation}
\begin{equation}
\mathcal{S}_{3,i,j}=r^2 (\mathcal{N}^2_{r,y}-\cos^2{\beta_r}),
\end{equation}
\end{subequations} 
where $[\mathcal{A}_{i,j},\mathcal{B}_{i,j},\mathcal{C}_{i,j}]$ is the unit direction vector of the sub-beam axis ${\rm{T}}_{i,j}$ and can be expressed as
\begin{subequations}
\begin{equation}
\mathcal{A}_{i,j}=\frac{\cos(i\beta)\cos(\vartheta_{i,j}+\vartheta_t)\cos(\alpha_t-\phi_{i,j})}{\cos{\vartheta_{i,j}}\cos{\phi_{i,j}}},
\end{equation}
\begin{equation}
\mathcal{B}_{i,j}=\frac{\cos(i\beta)\cos(\vartheta_{i,j}+\vartheta_t)\sin(\alpha_t-\phi_{i,j})}{\cos{\vartheta_{i,j}}\cos{\phi_{i,j}}},
\end{equation}
\begin{equation}
\mathcal{C}_{i,j}=\frac{\cos(i\beta)\sin(\vartheta_{i,j}+\vartheta_t)}{\cos{\vartheta_{i,j}}}.
\end{equation}
\end{subequations}
Besides, $\phi_{i,j}$ and $\vartheta_{i,j}$ can be given by
\begin{subequations}
\begin{equation}
\phi_{i,j}=\tan^{-1}\left[\frac{\cos{\vartheta_{i,j}}\tan(i\beta)\sin{\beta_j}}{\cos(\vartheta_{i,j}+\vartheta_t)}\right],
\end{equation}
\begin{equation}
\vartheta_{i,j}=\tan^{-1}[\tan(i\beta)\cos{\beta_j}],
\end{equation}
\end{subequations}
where 
\begin{subequations}
\begin{equation}
\beta_j=j\beta_i+2j\tan^{-1}\Upsilon(i),
\end{equation}
\begin{equation}
\beta_i=\frac{2\pi}{\nu_i}-2\tan^{-1}\Upsilon(i).
\end{equation}
\end{subequations}

\end{appendices}

\section*{Biographies}
\begin{IEEEbiography}[{\includegraphics[width=1in]{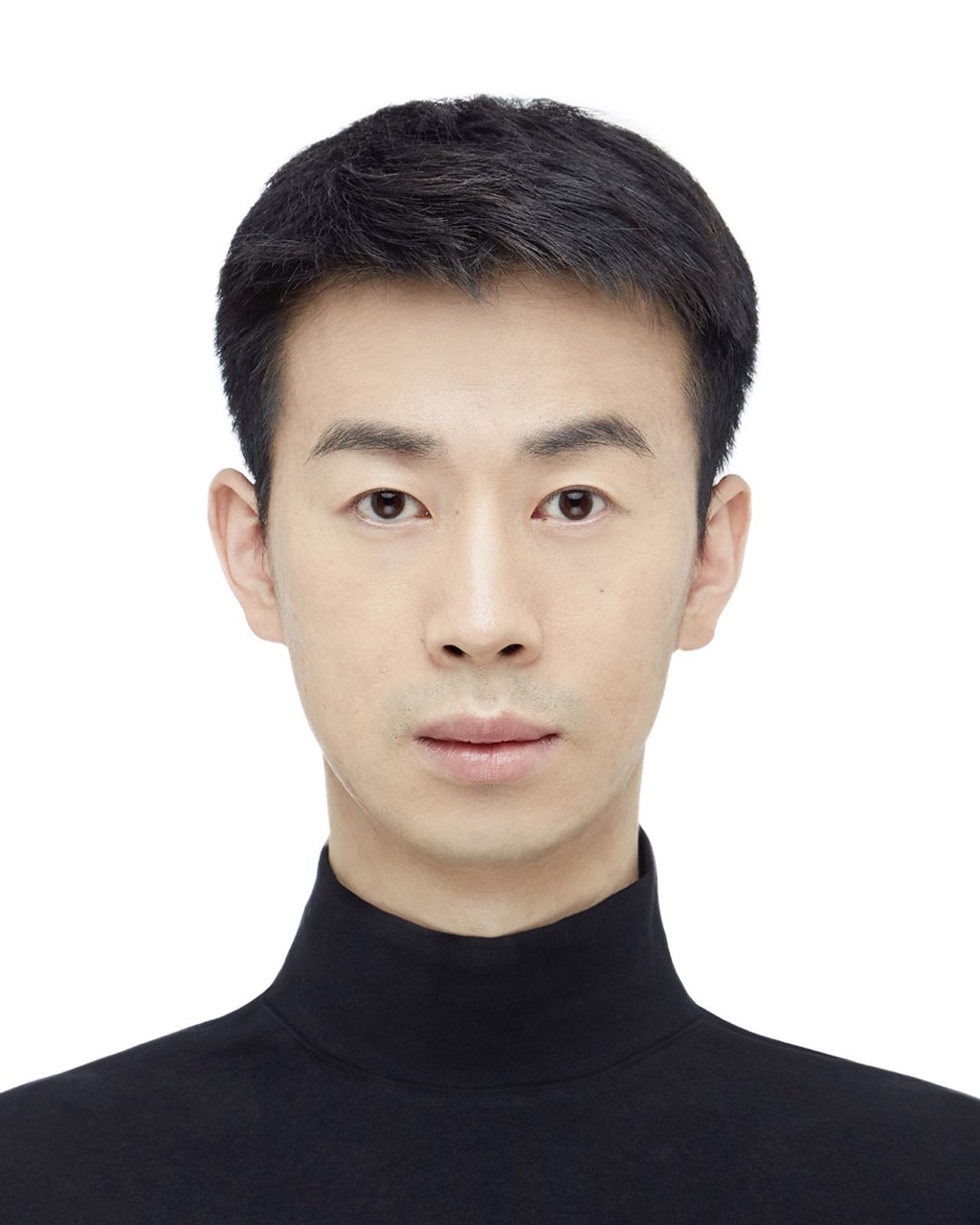}}]
	{Tianfeng Wu} received his B.S. degree from the College of Instrumentation and Electrical Engineering in Jilin University, Changchun, China, in 2015, and the double Master’s degree from the Department of Precision Instrument in Tsinghua University, Beijing, China, and the Faculty of Mechanical Engineering in RWTH Aachen University, Aachen, Germany, in 2019. He received a Deutscher Akademischer Austausch Dienst (DAAD) scholarship from 2015 to 2016. He is currently pursuing a Ph.D. degree with the Department of Electronic Engineering, Tsinghua University. His research interests include optical scattering communications and channel modeling.
\end{IEEEbiography}
~\\
\begin{IEEEbiography}[{\includegraphics[width=1in]{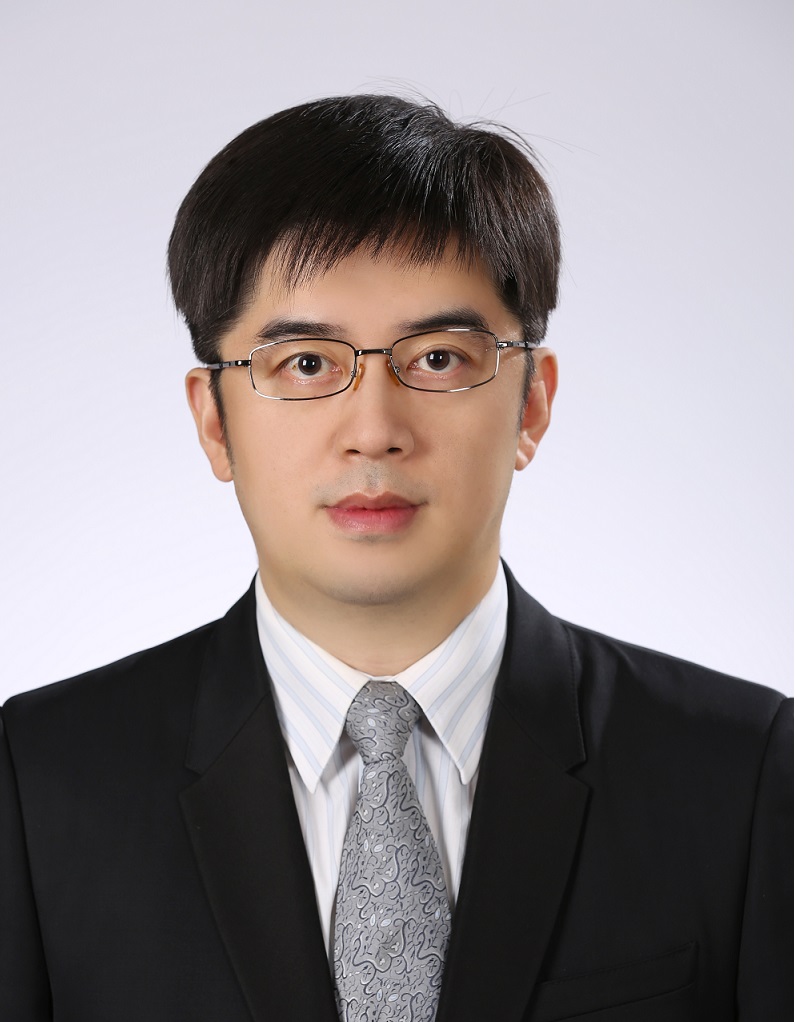}}]
	{Fang Yang} (M'11-SM'13) received the B.S.E. and Ph.D. degrees in electronic engineering from Tsinghua University, Beijing, China, in 2005 and 2009, respectively.
  He is currently an Associate Professor with the Department of Electronic Engineering, Tsinghua University. He has published over 180 peer-reviewed journal and conference papers. He holds over 50 Chinese patents and two PCT patents. His research interests are in the fields of power line communication, visible light communication, wireless communication, and digital television terrestrial broadcasting.
  Dr. Yang received the IEEE Scott Helt Memorial Award (Best Paper Award in the IEEE TRANSACTIONS ON BROADCASTING) in 2015. He is the Secretary General of Sub-Committee 25 of the China National Information Technology Standardization (SAC/TC28/SC25). He currently serves as an Associate Editor for the IEEE TRANSACTIONS ON GREEN COMMUNICATIONS AND NETWORKING and IET Optoelectronics. He is a Fellow of IET.
\end{IEEEbiography}
~\\
\begin{IEEEbiography}[{\includegraphics[width=1in]{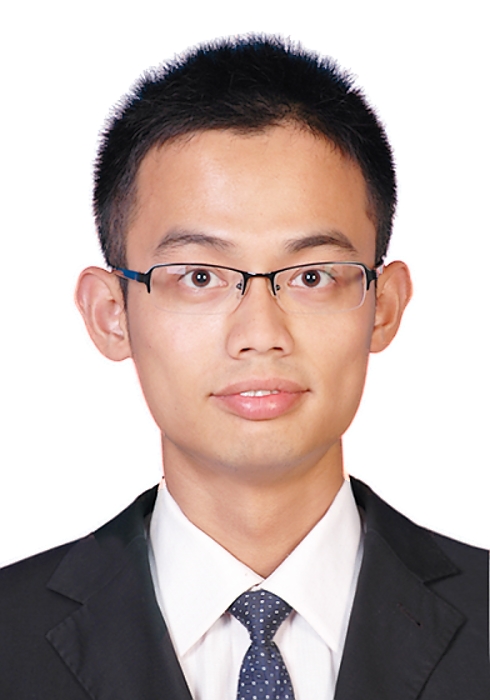}}]
	{Tian Cao} received the B.S. and M.S. degrees in telecommunication engineering from Xidian University, China, in 2014 and 2017, respectively, and the Ph.D. degree from the Department of Electronic Engineering, Tsinghua University, China, in 2023. He is currently working as a lecturer with the School of Telecommunications Engineering, Xidian University. His research interests include ultraviolet communications and visible light communications.
\end{IEEEbiography}
~\\
\begin{IEEEbiography}[{\includegraphics[width=1in]{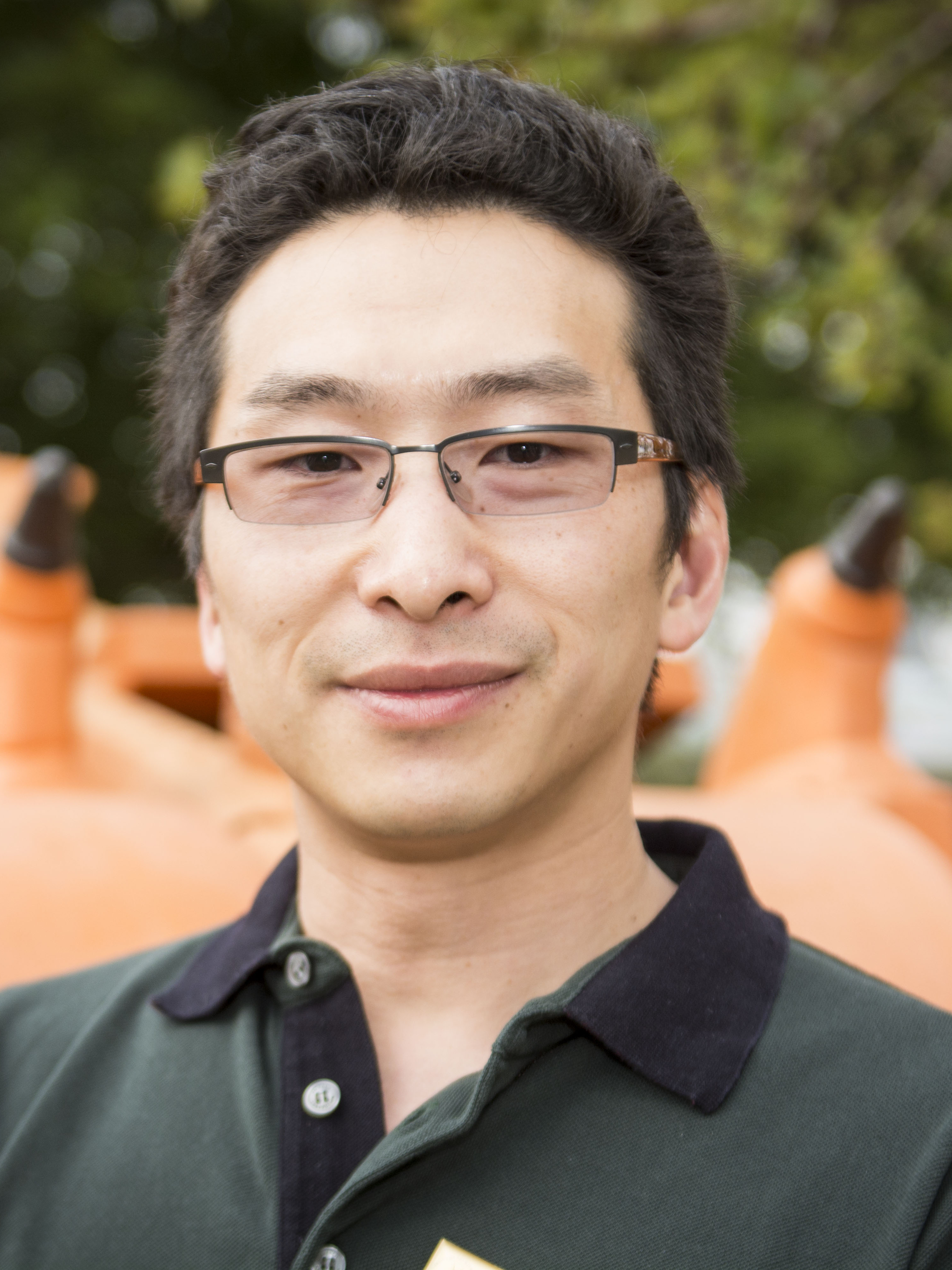}}]
	{Ling Cheng} (M'10-SM'15) received the degree B. Eng. Electronics and Information (cum laude) from Huazhong University of Science and Technology (HUST) in 1995, M. Ing. Electrical and Electronics (cum laude) in 2005, and D. Ing. Electrical and Electronics in 2011 from University of Johannesburg (UJ). His research interests are in Telecommunications and Artificial Intelligence. In 2010, he joined University of the Witwatersrand where he was promoted to Full Professor in 2019. He serves as the associate editor of three journals. He has published more than one hundred research papers in journals and conference proceedings. He has been a visiting professor at five universities and the principal advisor for over forty full research post-graduate students. He was awarded the Chancellor’s medals in 2005, 2019 and the National Research Foundation ratings in 2014, 2020. The IEEE ISPLC 2015 best student paper award was made to his Ph.D. student in Austin. He is a senior member of IEEE and the vice-chair of IEEE South African Information Theory Chapter.
\end{IEEEbiography}
~\\
\begin{IEEEbiography}[{\includegraphics[width=1in]{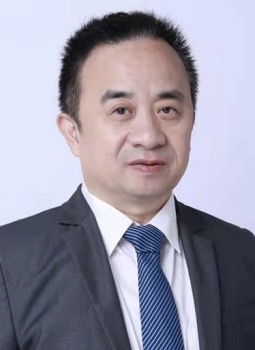}}]
	{Yupeng Chen} received both his B.Sc. and Ph.D. degrees from Tsinghua University, China. He is currently an Associate Professor at the Southern University of Science and Technology (SUSTech). His research focuses on micro-nano sensor devices, audio and video signal processing, and AI technology applications.
\end{IEEEbiography}
~\\
\begin{IEEEbiography}[{\includegraphics[width=1in]{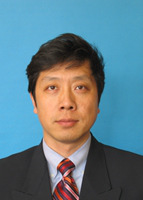}}]
	{Jian Song} (M'06-SM'10-F'16) received the B.Eng. and Ph.D. degrees in electrical engineering from Tsinghua University, Beijing, China, in 1990 and 1995, respectively.
  He is currently the Director of Tsinghua DTV Technology Research and Development Center, Tsinghua University. He has been working in quite different areas, including fiber-optic, satellite, and wireless communications, as well as the power-line communications. He holds two U.S. and more than 80 Chinese patents. He has published more than 300 peer-reviewed journal and conference papers and coauthored several books in the aforementioned areas. His current research interest is in the area of digital TV broadcasting, network convergence, and visible light communication networks.
  Dr. Song won several awards both international and domestic. He is also Fellow of IET, Chinese Institute of Electronics, and Chinese Institute of Communications.
\end{IEEEbiography}
~\\
\begin{IEEEbiography}[{\includegraphics[width=1in]{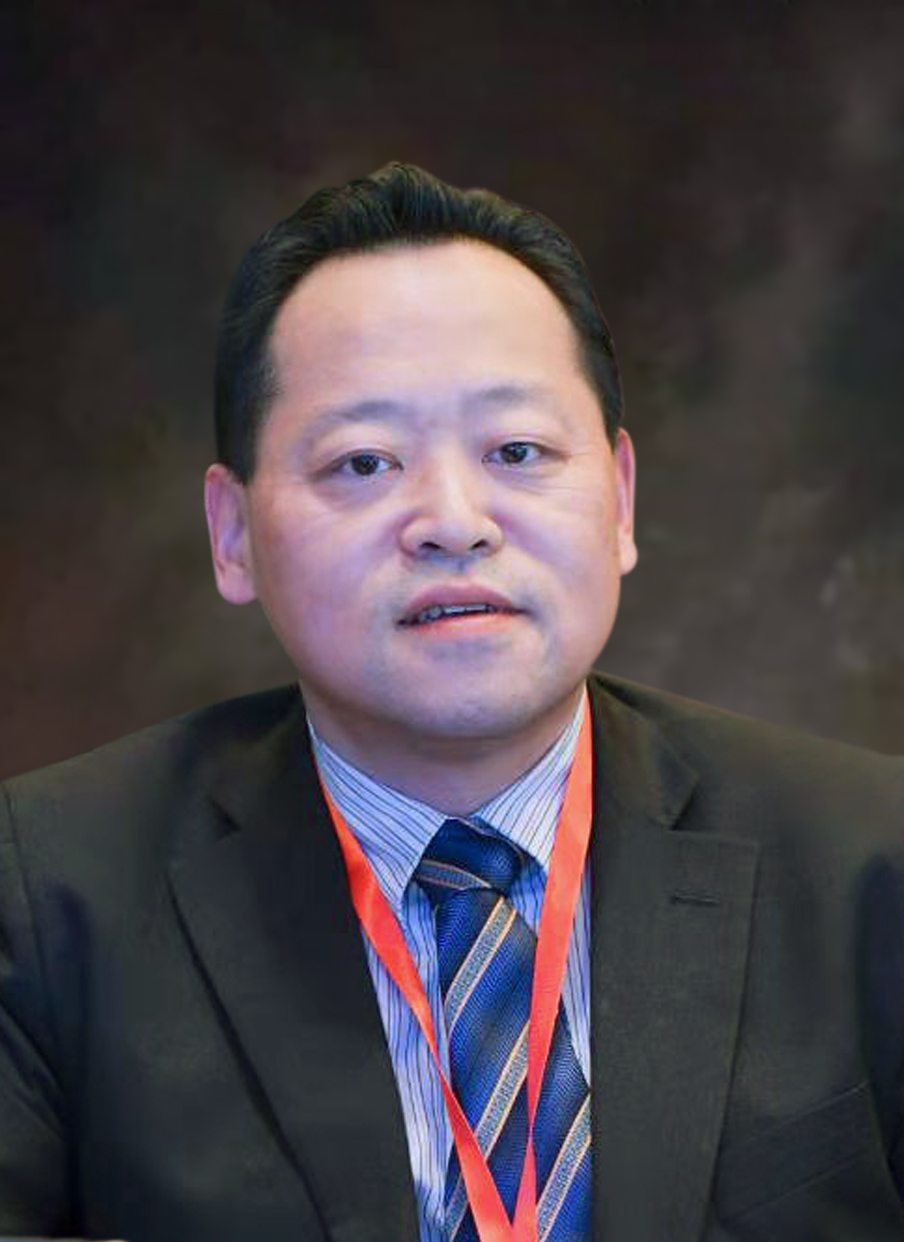}}]
	{Julian Cheng} (S’96–M’04–SM’13–F’23) received a B.Eng. degree (Hons.) in electrical engineering from the University of Victoria, Victoria, BC, Canada, in 1995, the M.Sc.(Eng.) degree in mathematics and engineering from Queen's University, Kingston, ON, Canada, in 1997, and a Ph.D. in electrical engineering from the University of Alberta, Edmonton, AB, Canada, in 2003. He is a Full Professor in the School of Engineering, Faculty of Applied Science, The University of British Columbia, Kelowna, BC, Canada. He was with Bell Northern Research and NORTEL Networks. His research interests include machine learning and deep learning for wireless communications, wireless optical technology, and quantum communications. He served as the President of the Canadian Society of Information Theory (2017-2021). He was the Co-Chair of the 12th Canadian Workshop on Information Theory in 2011, the 28th Biennial Symposium on Communications in 2016, and the General Co-Chair of the 2021 IEEE Communication Theory Workshop. He is the Chair of the Radio Communications Technical Committee of the IEEE Communications Society. He was a past Associate Editor of the IEEE TRANSACTIONS ON COMMUNICATIONS, the IEEE TRANSACTIONS ON WIRELESS COMMUNICATIONS, the IEEE COMMUNICATIONS LETTERS, and the IEEE ACCESS, as well as an Area Editor for the IEEE TRANSACTIONS ON COMMUNICATIONS. Dr. Cheng served as a Guest Editor for a Special Issue of the IEEE JOURNAL ON SELECTED AREAS IN COMMUNICATIONS on Optical Wireless Communications.
\end{IEEEbiography}
~\\
\begin{IEEEbiography}[{\includegraphics[width=1in]{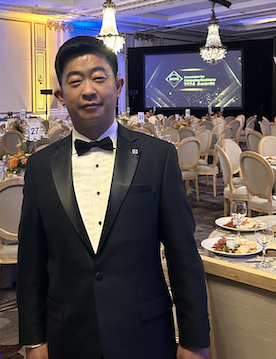}}]
	{Zhu Han} (S'01-M'04-SM'09-F'14) received the B.S. degree in electronic engineering from Tsinghua University, in 1997, and the M.S. and Ph.D. degrees in electrical and computer engineering from the University of Maryland, College Park, in 1999 and 2003, respectively. 

 From 2000 to 2002, he was an R\&D Engineer of JDSU, Germantown, Maryland. From 2003 to 2006, he was a Research Associate at the University of Maryland. From 2006 to 2008, he was an assistant professor at Boise State University, Idaho. Currently, he is a John and Rebecca Moores Professor in the Electrical and Computer Engineering Department as well as in the Computer Science Department at the University of Houston, Texas. Dr. Han’s main research targets on the novel game-theory related concepts critical to enabling efficient and distributive use of wireless networks with limited resources. His other research interests include wireless resource allocation and management, wireless communications and networking, quantum computing, data science, smart grid, carbon neutralization, security and privacy.  Dr. Han received an NSF Career Award in 2010, the Fred W. Ellersick Prize of the IEEE Communication Society in 2011, the EURASIP Best Paper Award for the Journal on Advances in Signal Processing in 2015, IEEE Leonard G. Abraham Prize in the field of Communications Systems (best paper award in IEEE JSAC) in 2016, IEEE Vehicular Technology Society 2022 Best Land Transportation Paper Award, and several best paper awards in IEEE conferences. Dr. Han was an IEEE Communications Society Distinguished Lecturer from 2015 to 2018 and ACM Distinguished Speaker from 2022 to 2025, AAAS fellow since 2019, and ACM Fellow since 2024. Dr. Han is a 1\% highly cited researcher since 2017 according to Web of Science. Dr. Han is also the winner of the 2021 IEEE Kiyo Tomiyasu Award (an IEEE Field Award), for outstanding early to mid-career contributions to technologies holding the promise of innovative applications, with the following citation: ``for contributions to game theory and distributed management of autonomous communication networks."
\end{IEEEbiography}
\end{document}